\documentclass[aps, prb, twocolumn, superscriptaddress, english]{revtex4-2}

\usepackage{graphicx}
\usepackage{cancel}
\usepackage{enumerate}
\usepackage{hyperref}
\hypersetup{
colorlinks=true,
citecolor=blue,
linkcolor=blue,
urlcolor=blue,
}
\usepackage{textcomp}
\usepackage{amsmath}
\usepackage{amssymb}
\usepackage{pgf}
\usepackage{soul}
\usepackage[english]{babel}
\usepackage[normalem]{ulem}
\usepackage{tikz}
\usetikzlibrary{arrows.meta,decorations.pathreplacing}
\usetikzlibrary{3d}
\usetikzlibrary{calc}
\usetikzlibrary{patterns}
\usetikzlibrary{fadings}
\usepackage{graphicx}
\usepackage{subcaption}
\usepackage{float}

\usepackage{orcidlink}

\def\ket#1{|#1\rangle}

\graphicspath{{./}{./Figure/}{./figures}}

\begin{document}
\title{Exploring ground state phase diagrams of Rydberg atom arrays\\ with finite projected entangled-pair states}

% \author{KS, ELW and MR}
\author{Konrad Schröder \orcidlink{0009-0007-4705-5147}}%
\affiliation{Institute for Theoretical Physics, University of Cologne, D-50937 K\"oln, Germany}
\affiliation{%
 Forschungszentrum J\"ulich  GmbH, Institute of Quantum Control, Peter Gr\"unberg Institut (PGI-8), 52425 J\"ulich, Germany
}%
\author{Erik Lennart Weerda
\orcidlink{0009-0006-8139-8648}}%
\affiliation{Institute for Theoretical Physics, University of Cologne, D-50937 K\"oln, Germany}
\author{Matteo Rizzi \orcidlink{0000-0002-8283-1005}}%
\affiliation{Institute for Theoretical Physics, University of Cologne, D-50937 K\"oln, Germany}
\affiliation{%
 Forschungszentrum J\"ulich  GmbH, Institute of Quantum Control, Peter Gr\"unberg Institut (PGI-8), 52425 J\"ulich, Germany
}%

\date{\today}

\begin{abstract}

Rydberg atom arrays (RAA) have been advocated as a promising avenue for the quantum simulation of different exotic quantum states. 
Numerical predictions for the ground state properties of this experimental platform can, however, be challenging in two-dimensional setups.
In particular, faithfully treating long-range interactions is often crucial for distinguishing the fate of competing phases, but it usually comes at considerable cost.
Projected entangled-pair states (PEPS), optimized via a sampling-based approach, have been shown to deal with such terms on the same ground as shorter-range ones. 
In this work, we employ these tools to investigate debated physical questions in different frustrated geometries of RAA, from the emergence of spin-glass physics on the Kagomé lattice to topological order beyond the PXP model on the Ruby lattice. By that, we contribute to establishing PEPS as well-suited tools for creating digital twins of RAA experiments.

\end{abstract}

\maketitle

\section{Introduction}
Rydberg atom arrays (RAAs) have emerged as a promising platform for both quantum simulation and quantum computation \cite{morgado_quantum_2021, wu_concise_2021, browaeys_many-body_2020}. The use of optical tweezers enables the preparation of arbitrary atomic geometries \cite{barredo_atom-by-atom_2016}, while Rabi driving and strong van-der-Waals interactions between Rydberg states provide access to many-body systems with highly programmable interactions. This has enabled the experimental exploration of strongly correlated quantum phases \cite{ebadi_quantum_2021, doi:10.1126/science.abi8794}.

A central objective of numerical simulations is to reliably predict the behavior observed in quantum simulator experiments. 
Such digital twins of experiments are essential both for benchmarking the performance of the simulator and for identifying the physical mechanisms underlying observed phenomena, and find relevant parameter regimes where interesting phenomena are expected to occur. 
For Rydberg atom arrays, this task is particularly challenging in two spatial dimensions, because the finite size and open boundary conditions of experimental systems can strongly influence the observed phases, while the long-range nature of the van-der-Waals interaction makes faithful numerical modeling more demanding than in conventional short-range lattice models. 
Consequently, numerical approaches that capture both the experimental geometry and the full interaction profile are crucial for establishing a reliable connection between theory and experiment \cite{orourke_entanglement_2023}. 
Thus, realistic numerical modeling is not only essential for establishing quantitative agreement with experiments, but also provides a route to investigate how robust fundamental many-body phenomena remain when idealized theoretical descriptions are extended towards experimentally relevant conditions.

Projected entangled-pair states (PEPS) \cite{verstraete_renormalization_2004, orus_practical_2014} provide a natural framework for addressing these questions in two dimensions. 
PEPS emerged as the natural tensor-network generalization of matrix product states to higher dimensions and provide a variational class naturally compatible with area-law entanglement scaling \cite{eisert_colloquium_2010, verstraete_criticality_2006}. 
In this respect, they are more natural and expressive than their quasi-1D counterparts, and they should avoid mistaking the stability of certain phases.
However, their application to large finite systems has historically been limited by the computational complexity of tensor-network contractions and optimization \cite{lubasch_algorithms_2014, scheb_finite_2023}. Recently, sampling-based approaches to PEPS optimization have enabled efficient ground-state searches on finite lattices~\cite{PhysRevB.111.195120, chan_efficient_2026, neuscamman_optimizing_2012, wang_monte_2011, liu_gapless_2018, liu_accurate_2021, liu_gapless_2022, liu_accurate_2025, dong_efficient_2025, Sandvik2007}. In particular, this approach is well suited for Rydberg atom arrays, as long-range interactions diagonal in the computational basis can be treated directly without introducing uncontrolled interaction truncations \cite{PhysRevB.111.195120, chan_efficient_2026}.

In this work, we employ finite PEPS to investigate the ground-state phase diagrams of Rydberg atom arrays on several frustrated geometries. These systems have attracted considerable interest due to the emergence of competing ordered phases and proposals for exotic quantum states. On the Kagomé lattice, previous studies have reached conflicting conclusions regarding the existence of an intermediate spin-glass phase \cite{PhysRevLett.130.206501, hibat-allah_recurrent_2025}. On the Ruby lattice, a $\mathbb{Z}_2$ topologically ordered phase has been proposed in the constrained PXP limit \cite{PhysRevX.11.031005}, raising the question of whether this phase can remain stable when realistic long-range interactions are restored \cite{doi:10.1126/science.abi8794}.

We first benchmark our approach on the triangular lattice and identify the resulting density-wave ordered phases and disordered regimes in Sec.~\ref{sec:results_triangular}. 
We then revisit the Kagomé lattice using open boundary conditions and the full long-range interaction, finding no evidence for a stable spin-glass phase in Sec.~\ref{sec:results_kagome}. 
Instead, the intermediate regime is consistent with a homogeneous disordered phase. 
Finally, we study the Ruby-lattice model and characterize the previously proposed topologically ordered phase using non-local Fredenhagen-Marcu order parameters and the topological contribution to the entanglement entropy in Sec~\ref{sec:results_ruby}. 
While we reproduce the $\mathbb{Z}_2$ spin liquid in the PXP limit, we find that realistic long-range interactions destabilize the topological phase, as soon as they are considered beyond a few neighbours.
Our results demonstrate the importance of boundary conditions and interaction truncations in the interpretation of Rydberg atom array phase diagrams and highlight finite PEPS as a powerful approach for creating reliable digital twins of quantum simulation experiments.
\section{Model and methods}

\subsection{Model of Rydberg atom array experiments}
In Rydberg atom array experiments, the use of optical tweezers allows the individual spatial rearrangement of the neutral atoms \cite{barredo_atom-by-atom_2016, browaeys_many-body_2020}. On the one hand, this has been used for the realization of different ordered or disordered configurations of the atoms. By changing the interatomic distance within a given configuration, however, this further allows tuning the strength of the van-der-Waals interaction between the atoms in the Rydberg state. 
In addition to the optical tweezers, these experiments use near-resonant laser fields that utilize a two-photon process to achieve a driving with detuning between the ground and the Rydberg state of the neutral atoms.

To describe such an experimental setup, we use the following Hamiltonian,
\begin{equation}
        \hat{H} = \frac{1}{2}\sum_i  \hat{\sigma}^x_i - \Delta \sum_i \hat{n}_i + \sum_{i \neq j} \frac{1}{((\textbf{r}_i - \textbf{r}_j)/R_b)^6}\hat{n}_i \hat{n}_j.
\label{eq:rydberg_hamiltonian}
\end{equation}
The basis of the local two-dimensional Hilbert spaces is spanned by the ground and Rydberg states of the atoms ($\ket{g},\ket{r}$). The Rabi frequency, characterizing the strength of the driving between ground and Rydberg states, is used to set the energy scale of the Hamiltonian. The detuning $\Delta$ allows us to energetically favor Rydberg states, while the blockade radius $R_b$ is used to characterize the strength of the van-der-Waals interaction between Rydberg states. 

This Hamiltonian hosts a plethora of ordered and disordered phases as ground states \cite{ebadi_quantum_2021, orourke_entanglement_2023} and has been used to propose the ground-state realization of topologically ordered states \cite{PhysRevX.11.031005, verresen_unifying_2022, doi:10.1126/science.abi8794} on the ruby lattice.

%%%%%%%%%%%%%%%%%%%%%%%%%%%%%%%%%%%%%%%%%%%%%%%%%%%%%%%%%%%%%%%%%%%%%%%%%%%%%%%%%%%%%%%%%%%%%%%%
\subsection{Finite projected entangled-pair states}
\label{sec:finitePEPS}
To analyse the phases of the RAA model on various lattices, we use projected entangled pair states \cite{verstraete_renormalization_2004}. 
These states are the paradigmatic examples of quantum states represented as tensor networks in two spatial dimensions and fulfill the boundary law of entanglement entropy \cite{verstraete_criticality_2006, eisert_colloquium_2010}. 
The fundamental building blocks of finite PEPS are local tensors, which are defined on each lattice site,
\begin{equation} 
T[x,y] = \begin{gathered}
        \includegraphics[height=2cm]{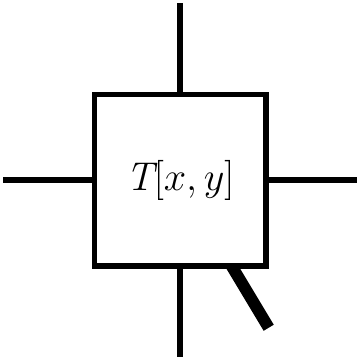}
\end{gathered}\;.\end{equation}
Every local tensor has a single \textit{physical} index (bold line), labeling a basis of its local physical Hilbert space of dimension $d$, as well as a set of \textit{virtual} indices (horizontal and vertical lines) of dimension $D$ that connect to the local tensors on neighboring sites. 
Within the PEPS ansatz, we express the coefficient tensor of the many-body state as the contraction of all virtual indices, illustrated as connected legs, of the local tensors,
\begin{equation}
\label{PsiS}
\begin{split}
    \ket{\Psi} = \sum_{\{\textbf{S}\}} &\Psi(\textbf{S}) \ket{\textbf{S}}, \quad \ket{\textbf{S}}=\ket{s_{(1,1)}\dots s_{(L_x,L_y)}}\\
    \Psi(\textbf{S}) &= \begin{gathered}
        \includegraphics[height=3.8cm]{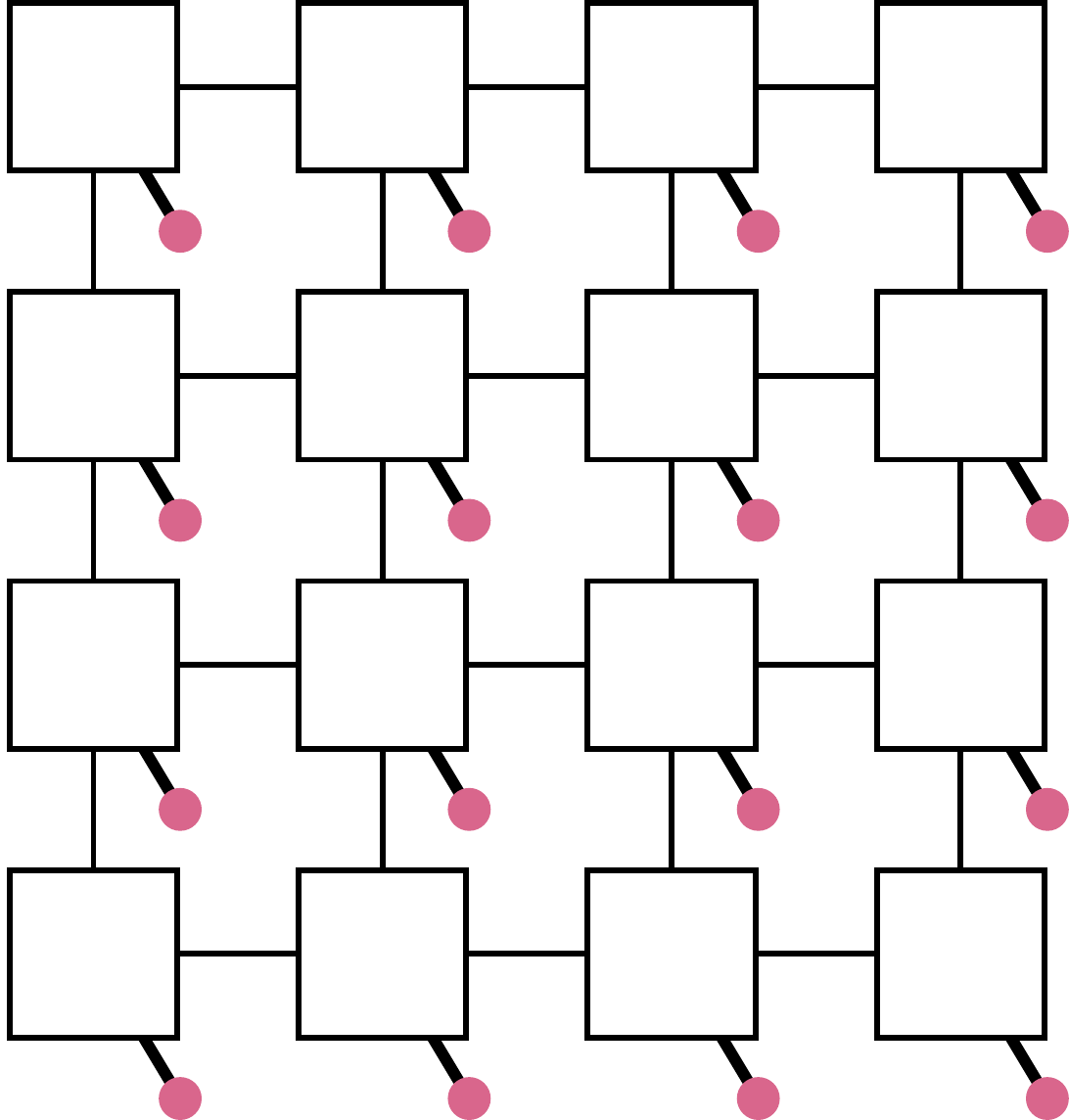}
    \end{gathered}
\end{split}
\end{equation}
Here, $\textbf{S}$ denotes a single many-body configuration, and the red dots signify that we fixed the index at position $(x,y)$ to a certain value $s_{(x,y)}$.

The above illustrates a finite PEPS defined on a square lattice. To study non-square lattices, we will employ an appropriate mapping from the lattice in question to the square lattice. 
We will further discuss these mappings in the relevant sections of Sec.~\ref{sec:results}.

While several numerical approaches for the finite PEPS have been formulated since their inception, we employ a direct sampling-based approach \cite{wang_monte_2011, vieijra_direct_2021} for ground state search using the time-dependent variational principle \cite{sorella_weak_2007, haegeman_time-dependent_2011-1, carleo_solving_2017, vanderstraeten_tangent-space_2019, PhysRevB.111.195120}. 
This approach has recently been successfully applied to several important systems ranging from frustrated spin systems, including chiral spin liquids, to correlated fermions and Rydberg atom arrays \cite{liu_gapless_2018, liu_accurate_2021, liu_gapless_2022, liu_accurate_2025, PhysRevB.111.195120}. Additionally, it was recently applied to real-time evolution \cite{Wu2026}.

For the study of the Rydberg atom arrays, however, the approach offers an additional unique advantage. For long-range interactions that are diagonal in the computational basis, which is the case in Eq.~\eqref{eq:rydberg_hamiltonian}, the sampling-based approach allows for the efficient treatment of these long-range terms without the need for truncations at some finite distance. 
This has been shown to be critical to correctly identify the ground state phases of RAA Hamiltonians \cite{orourke_entanglement_2023}.

Before proceeding, let us notice that the use of PEPS is well justified also in the presence of algebraically decaying interactions.
Indeed, a weaker notion of locality is expected to hold for interactions decaying with a large enough power of the distance: It has been shown in such systems that the spread of correlations can still show a light-cone structure and follow generalized Lieb-Robinson bounds \cite{PhysRevLett.111.207202}, as explicitly discussed in the context of Rydberg atom arrays \cite{foss-feig_nearly_2015}.
Consequently, the boundary law for entanglement entropy still holds~\cite{samajdar_complex_2020, orourke_entanglement_2023}, and an efficient description in terms of PEPS is amenable.
\section{Results}
\label{sec:results}

In this section, we apply our PEPS approach to triangular, Kagomé, and Ruby Rydberg lattices. We first benchmark the method on the triangular lattice and determine some of its different phases alongside a rough sketch of the phase diagram. We then revisit the debated intermediate regime of the Kagomé lattice, where previous studies reported conflicting evidence for a spin-glass phase \cite{PhysRevLett.130.206501,hibat-allah_recurrent_2025}. Finally, we investigate the proposed $\mathbb{Z}_2$ topological phase of the Ruby-lattice PXP model \cite{PhysRevX.11.031005} and study its stability in the presence of realistic long-range interactions on a system with open boundary conditions.

%%%%%%%%%%%%%%%%%%%%%%%%%%%%%%%%%%%%%%%%%%%%%%%%%%%%%%%%%%%
\subsection{triangular lattice} \label{sec:results_triangular}

The triangular lattice is particularly interesting in comparison to the more extensively studied square lattice. Importantly, its larger coordination number results in a denser packing of atoms, such that each Rydberg excitation affects a greater number of neighboring sites. Consequently, the influence of the blockade radius $R_b$ on the phase diagram is expected to be more pronounced.

In order to apply our algorithm, we employ the one-to-one mapping of the triangular lattice onto the square lattice shown in Appendix~\ref{app:triangularmapping}, as is standard when studying non-square lattices \cite{naumann_introduction_2024, he_itinerant_2023}.

To characterize the different phases of the triangular-lattice model, we perform finite-PEPS calculations on finite triangular lattices with open boundary conditions and bond dimension $D = 4$. 
Depending on the expected ordering pattern, lattice sizes between $22 \times 22$ and $25 \times 25$ are employed to accommodate the corresponding unit cells and minimize finite-size effects associated with incommensurate boundaries. 
Selected calculations were repeated at $D = 5$, at small additional computational cost, yielding consistent results and indicating convergence.

To characterize the different phases, we compute the structure factor,
\begin{equation}
	S(\mathbf{k}) = \sum_{j,k} e^{i \mathbf{k} \cdot \mathbf{r}_{jk}} \langle \hat{n}_j \hat{n}_k \rangle,
    \label{eq:sf}
\end{equation}
where $\mathbf{r}_{jk} = \mathbf{r}_{k} - \mathbf{r}_{j}$ is the vector connecting the two sites in the real space lattice. The structure factor measures density–density correlations in momentum space and reveals how excitations are spatially arranged. Sharp peaks in $S(\mathbf{k})$ at specific wave vectors indicate the emergence of long-range crystalline order with characteristic periodicity, while a flat or diffuse spectrum is typical of a disordered phase. 

\begin{figure}[tbp] %{figure*}
    \centering
		\begin{subfigure}[b]{0.48\columnwidth} %{0.24\textwidth}
			\centering
			\includegraphics[width=\textwidth]{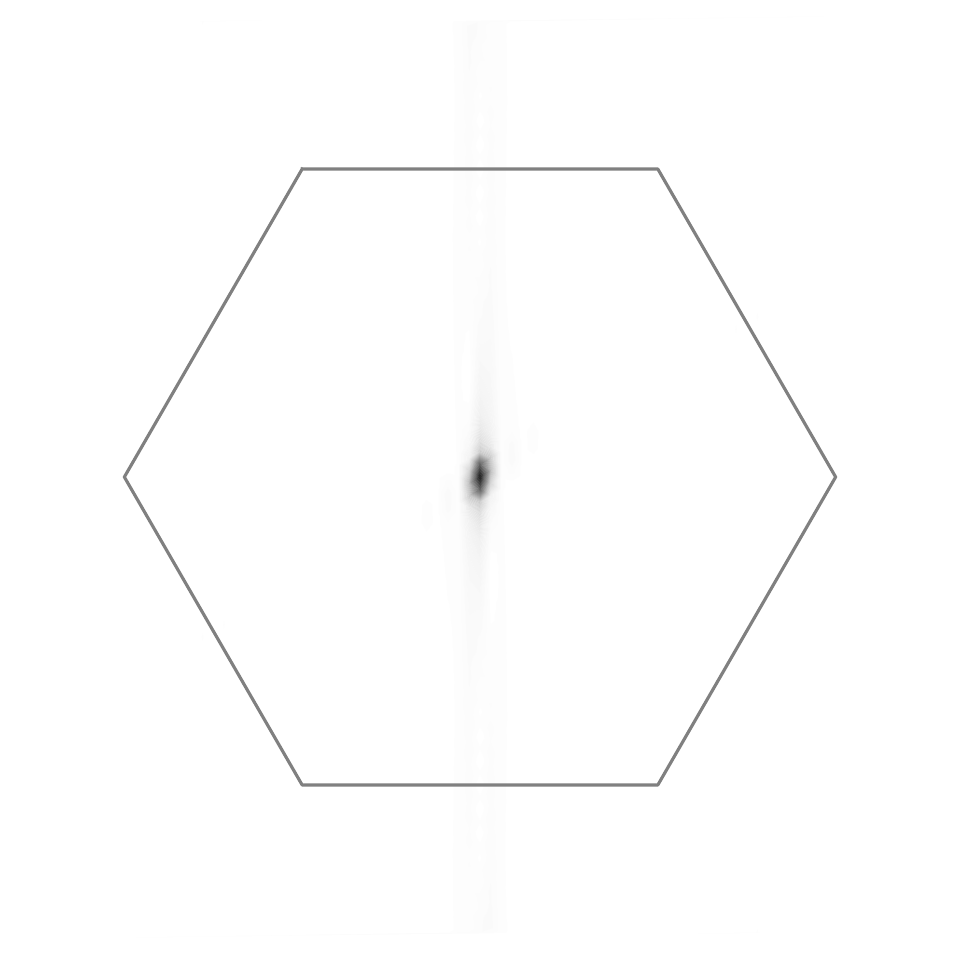}
			\caption{Disordered}
			\label{fig:sf-disordered}
		\end{subfigure}
		\hfill
		\begin{subfigure}[b]{0.48\columnwidth} %{0.24\textwidth}
			\centering
			\includegraphics[width=\textwidth]{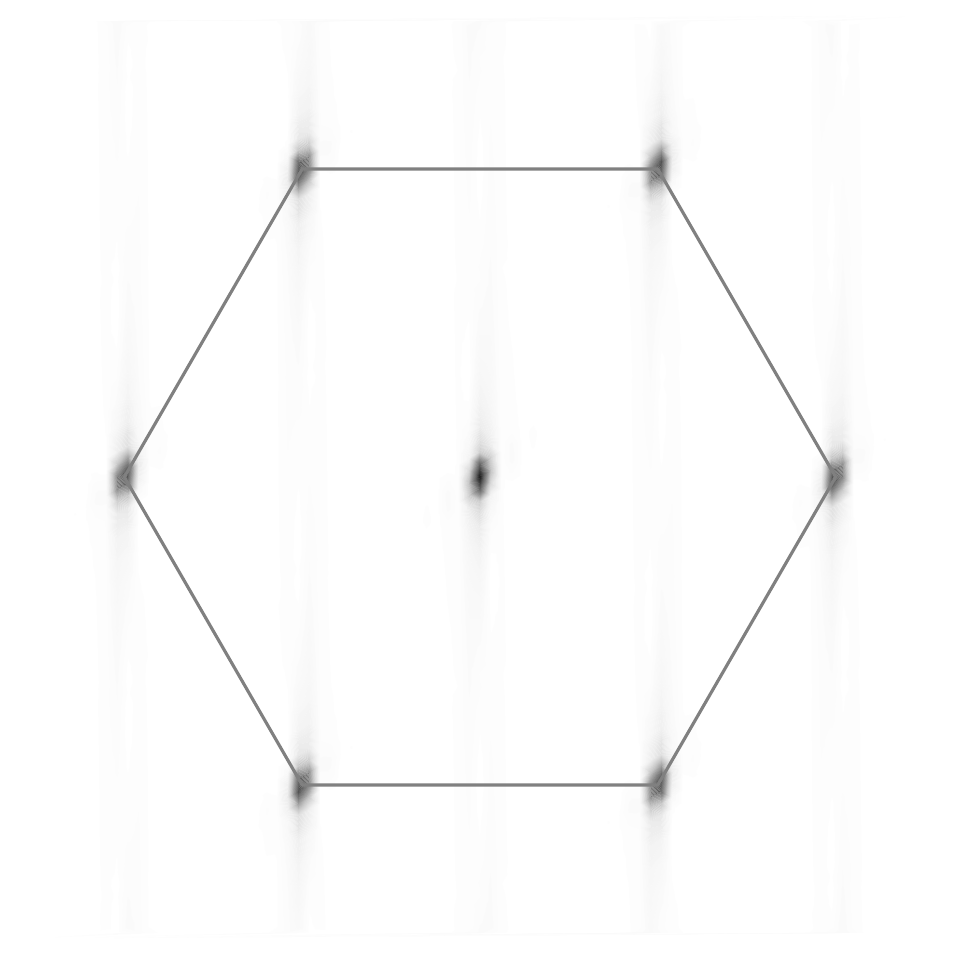}
			\caption{$\frac{1}{3}$-filling}
			\label{fig:sf-1over3}
		\end{subfigure}
  \\  \vspace{3mm} %		\hfill
        \begin{subfigure}[b]{0.48\columnwidth} %{0.24\textwidth}
			\centering
			\includegraphics[width=\textwidth]{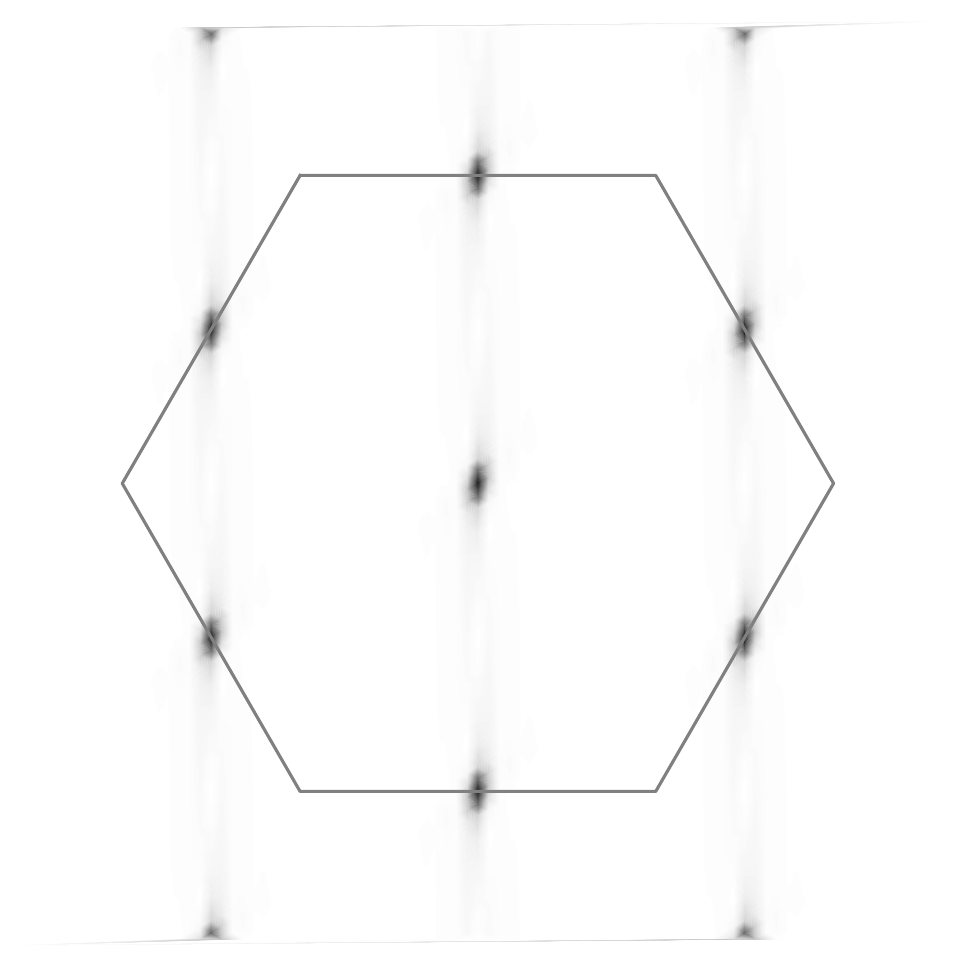}
			\caption{$\frac{1}{4}$-filling}
			\label{fig:sf-1over4}
		\end{subfigure}
		\hfill
         \begin{subfigure}[b]{0.48\columnwidth} %{0.24\textwidth}
			\centering
			\includegraphics[width=\textwidth]{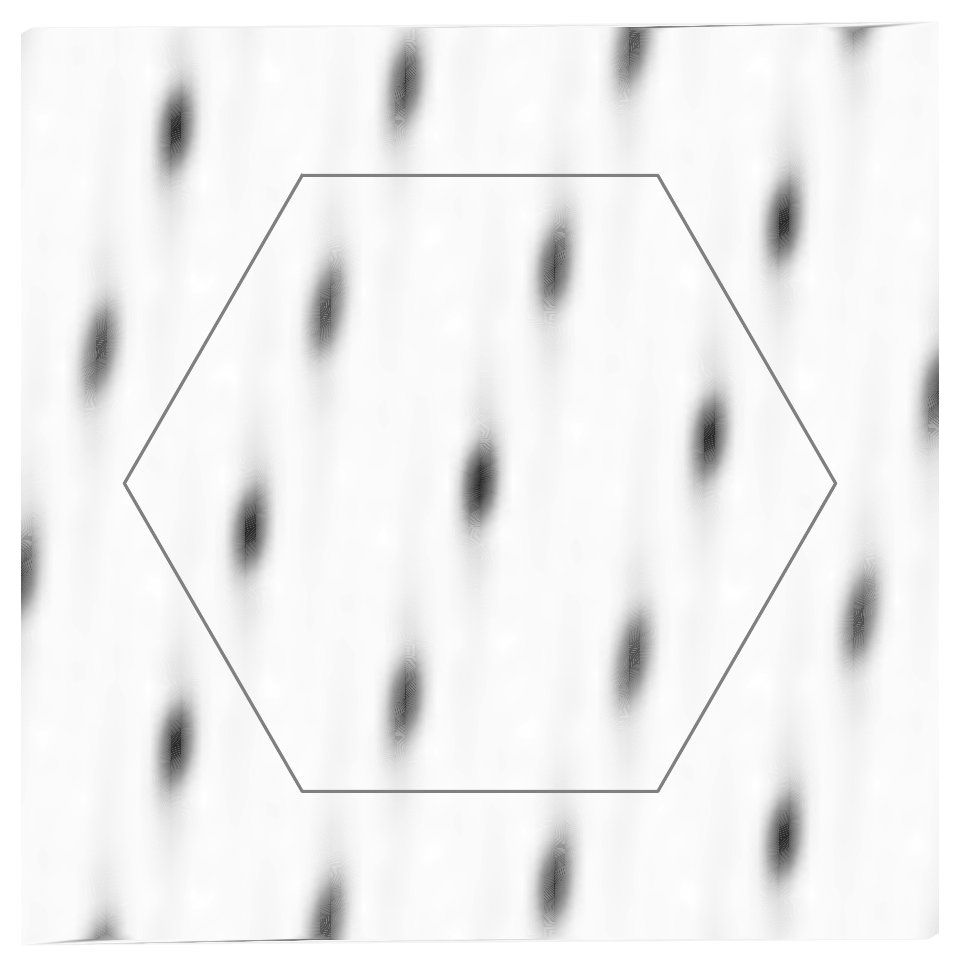}
			\caption{$\frac{1}{7}$-filling}
			\label{fig:sf-1over7}
		\end{subfigure}
		\caption
		{\small	Structure factor $S(\mathbf{k})$ for different phases of the triangular lattice Rydberg model. The locations in phase space are (a): $\Delta = 1.5$, $R_b = 1.8$, (b): $\Delta = 2$, $R_b = 1.2$, (c): $\Delta = 3$, $R_b = 1.6$, (d): $\Delta = 3.5$, $R_b = 2.1$. Ordered phases exhibit sharp Bragg peaks at characteristic wave vectors consistent with the underlying density-wave order, while the disordered phase shows a diffuse distribution without pronounced peak structure. With increasing blockade radius $R_b$, the dominant ordering wave vectors shift towards smaller momenta, reflecting progressively more dilute crystalline configurations. The structure factors for the disordered and the $\frac{1}{7}$-filling phase are evaluated in the bulk by excluding boundary sites to suppress finite-size edge effects arising from open boundary conditions.}
		\label{fig:sf-triangular}
\end{figure} %{figure*}

Since open boundary conditions energetically favor excitations near the system's edges, boundary effects can distort the measured correlations. To isolate the bulk behaviour, we therefore introduce a boundary-layer cutoff $b$, which specifies the number of outer atomic layers excluded from the structure-factor calculation. Here, $b=0$ corresponds to the full system, while increasing $b$ progressively removes additional boundary layers from the analysis.

In the explored parameter regime of $\Delta \in [1.0, 4.0]$ and $R_b \in [1.0, 2.2]$, we identify four different phases. 
These phases are in qualitative agreement with the phase structure reported for the triangular-lattice RAA model in previous studies~\cite{li_quantum_2022}.
Representative structure factors for each of them are shown in Fig.~\ref{fig:sf-triangular}. 
The displayed parameter points are chosen well inside each phase such that the characteristic ordering pattern is clearly resolved. 
The ordered phases exhibit sharp Bragg peaks whose positions reflect the underlying density-wave pattern, while the disordered phase shows no pronounced peak structure, consistent with the absence of long-range crystalline order. 
For all ordered phases, the Bragg peaks respect the sixfold rotational symmetry of the triangular lattice, indicating that the coarse-graining to a square lattice for computational purposes did not influence the physics.
Furthermore, increasing the blockade radius $R_b$ shifts the dominant ordering wave vectors towards smaller momenta, reflecting the increasing real-space separation between Rydberg excitations.

In Appendix~\ref{app:triangular}, we also map a qualitative phase diagram for the triangular-lattice RAA model, finding a phase structure in qualitative agreement with Ref.~\cite{li_quantum_2022}.

%%%%%%%%%%%%%%%%%%%%%%%%%%%%%%%%%%%%%%%%%%%%%%%%%%%%%%%%%%%%%%%%%%%%%%%%%%%%%%%%%%%%%%%%%%%%%%%%%%%%%%%%
\subsection{Kagomé lattice and the spin glass}
\label{sec:results_kagome}

The Kagomé Rydberg model has recently attracted considerable attention because of the debated existence of an intermediate glassy phase separating two distinct ordered phases. 
While quantum Monte Carlo calculations have reported evidence for such a spin-glass regime~\cite{PhysRevLett.130.206501}, more recent neural-network quantum state simulations found no indication of glassy ordering~\cite{hibat-allah_recurrent_2025}. 
While these studies reach different conclusions, they share two important approximations: both truncate the long-range Rydberg interactions and employ periodic boundary conditions. 
The former is particularly relevant, as interaction truncation is known to qualitatively modify the ground-state phase diagram \cite{orourke_entanglement_2023}, while the latter differs from the finite open-boundary geometries realized in current experiments. 

In this work, we revisit this question using our sampling-based PEPS ansatz, which naturally implements open boundary conditions and retains the full long-range Rydberg interaction. 
We coarse-grain three Kagomé sites into a single effective site, yielding a triangular lattice with local Hilbert-space dimension $d=2^3=8$; see App.~\ref{app:kagome}.
The resulting triangular lattice is mapped onto the square-lattice geometry as in the previous Section~\ref{sec:results_triangular}; see also App.~\ref{app:triangularmapping}.

\begin{figure}[t]
    \centering
		\begin{subfigure}[b]{0.23\textwidth}
			\centering
			\includegraphics[width=\textwidth]{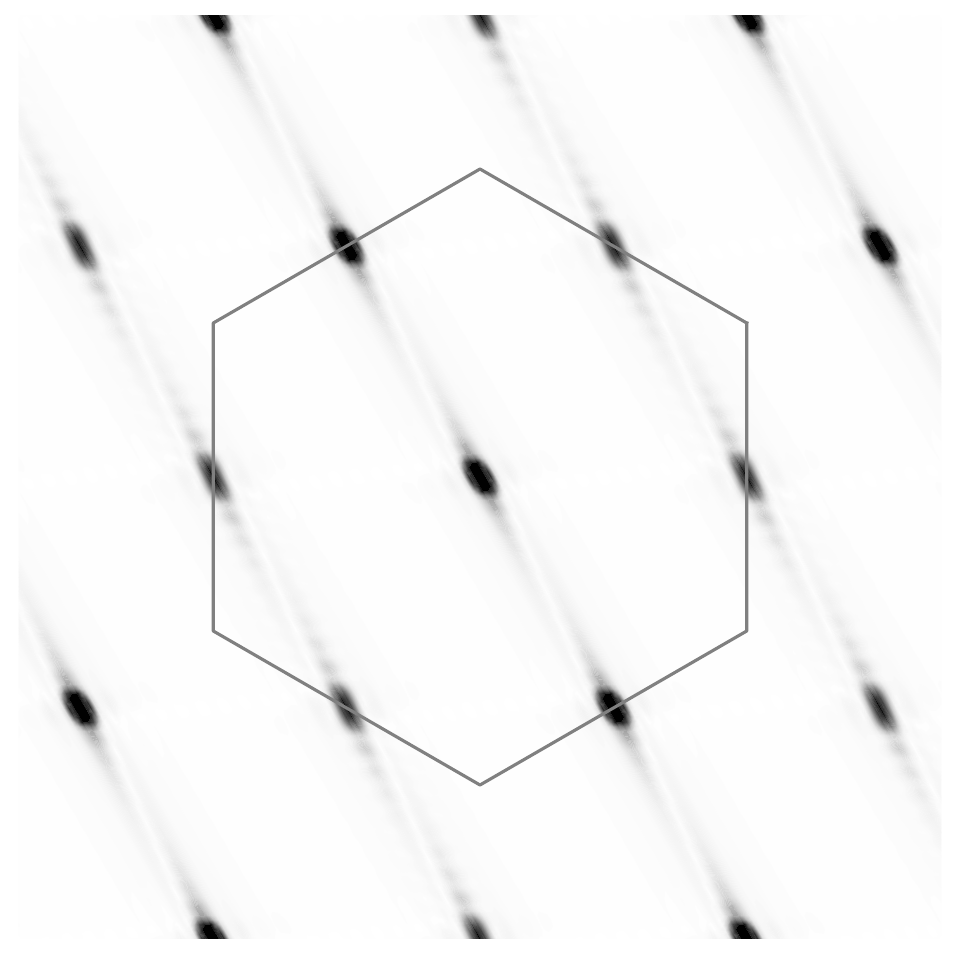}
			\caption{Nematic}
			\label{fig:sf-nematic}

		\end{subfigure}
		\hfill
		\begin{subfigure}[b]{0.23\textwidth}
			\centering
			\includegraphics[width=\textwidth]{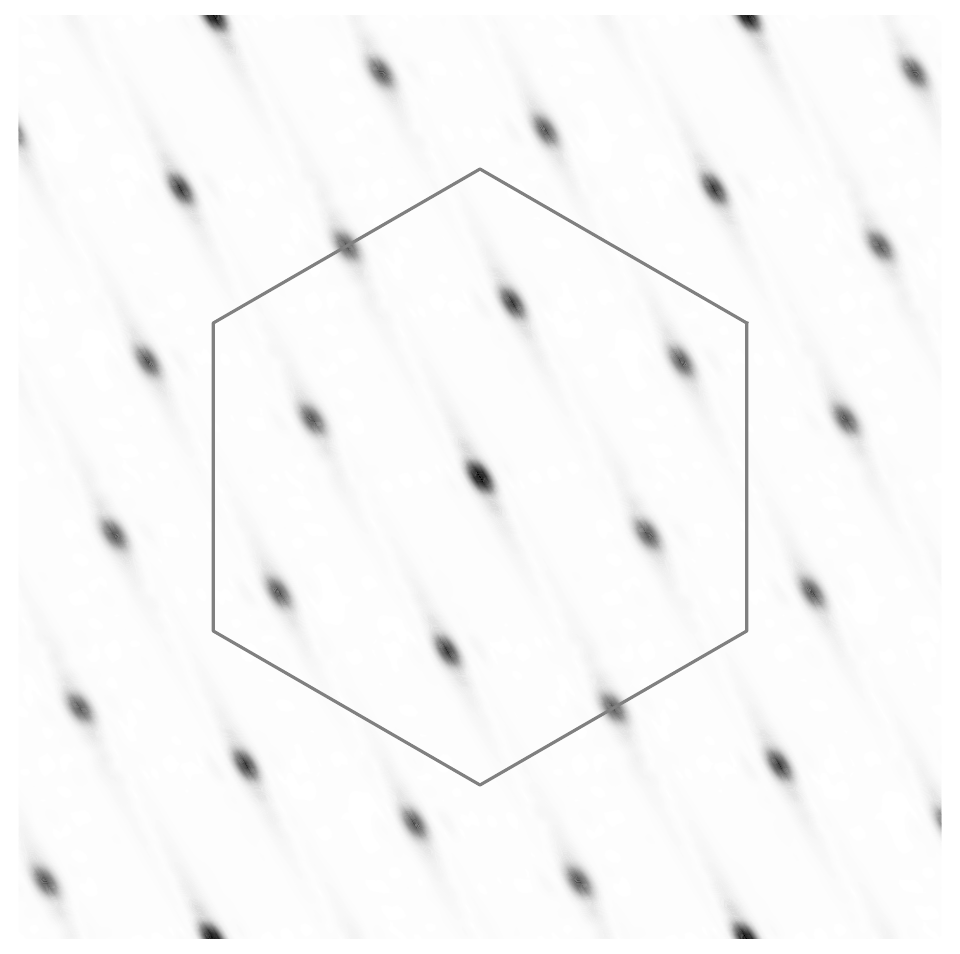}
			\caption{Staggered}
			\label{fig:sf-staggered}

		\end{subfigure}
		\caption
		{\small	Structure factor $S(\mathbf{k})$ for representative points deep within the ordered regions of the Kagomé Rydberg phase diagram. a) Nematic phase at $\Delta = 3.3$, $R_b = 1.6$. b) Staggered phase at $\Delta = 3.3$, $R_b = 2.3$. In both cases, sharp Bragg peaks appear at the expected wave vectors, confirming that our simulations correctly reproduce the ordered phases at the corresponding locations in parameter space.}
		\label{fig:sf-kagome}
\end{figure}

Here, we optimized a $12\times12$ array of coarse-grained sites, each representing three Kagomé lattice sites, corresponding to a total of $432$ physical atoms, using a PEPS with bond dimension $D=4$.
We first verify that the method reproduces the established nematic and staggered phases by computing the structure factor for parameters deep within the respective regions of the phase diagram see Fig.~\ref{fig:sf-kagome}, and confirm agreement with the results reported in \cite{doi:10.1073/pnas.2015785118}. 
In the intermediate regime, we observe no Bragg peaks, leaving two possible scenarios: a homogeneous disordered phase or a glassy state characterized by an irregular but frozen excitation pattern. 

 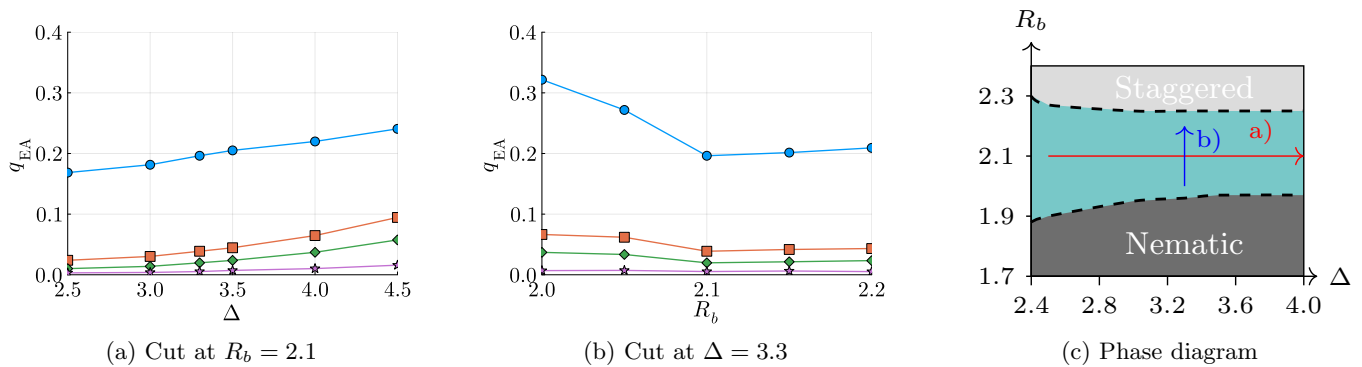
\begin{figure*}[htbp]
    \centering
		\begin{subfigure}[b]{0.3\textwidth}
			\centering
			\includegraphics[width=\textwidth]{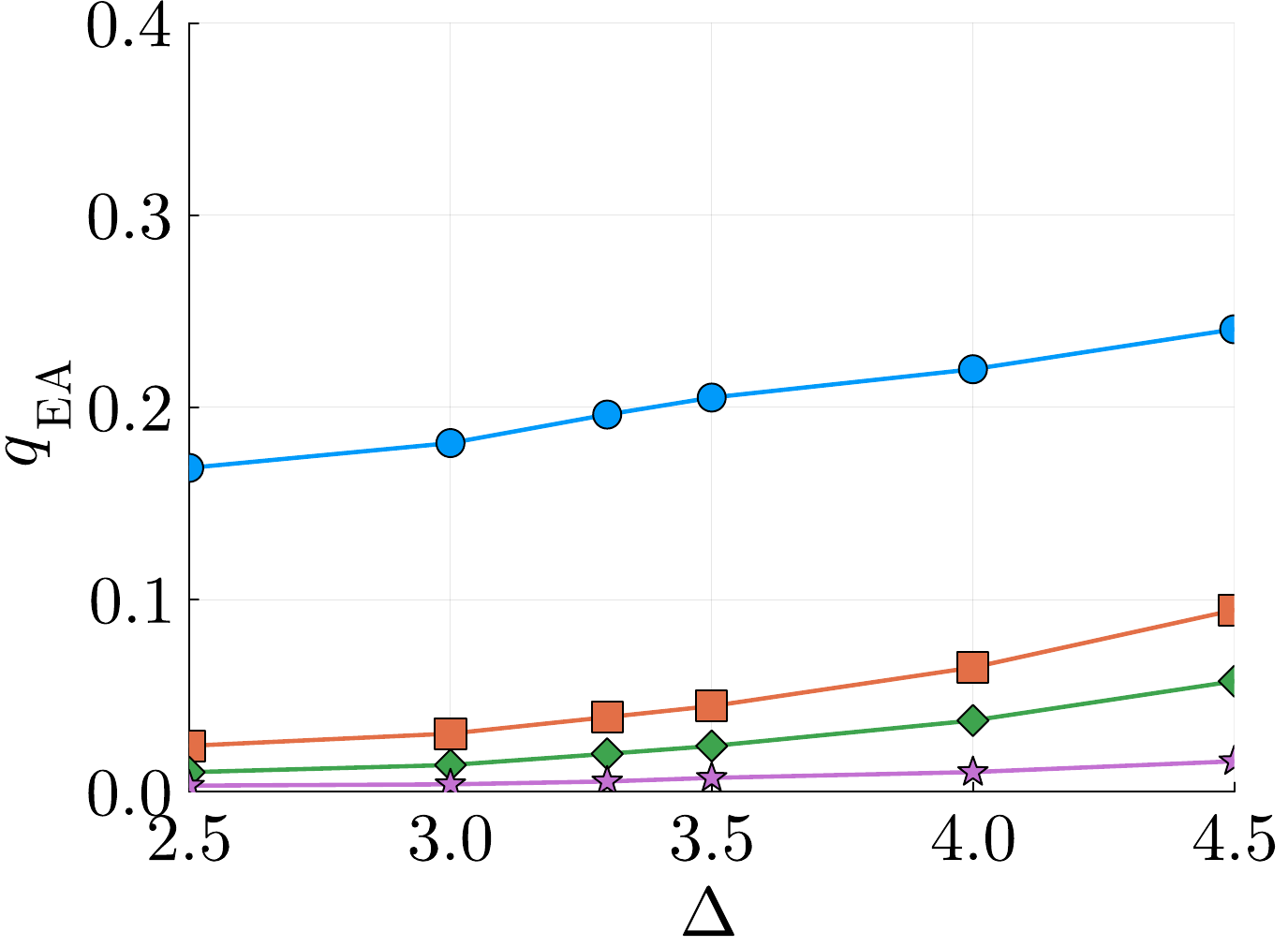}
			\caption{Cut at $R_b=2.1$}
			\label{fig:EA-cutB}

		\end{subfigure}
		\hfill
		\begin{subfigure}[b]{0.3\textwidth}
		\centering
			\includegraphics[width=\textwidth]{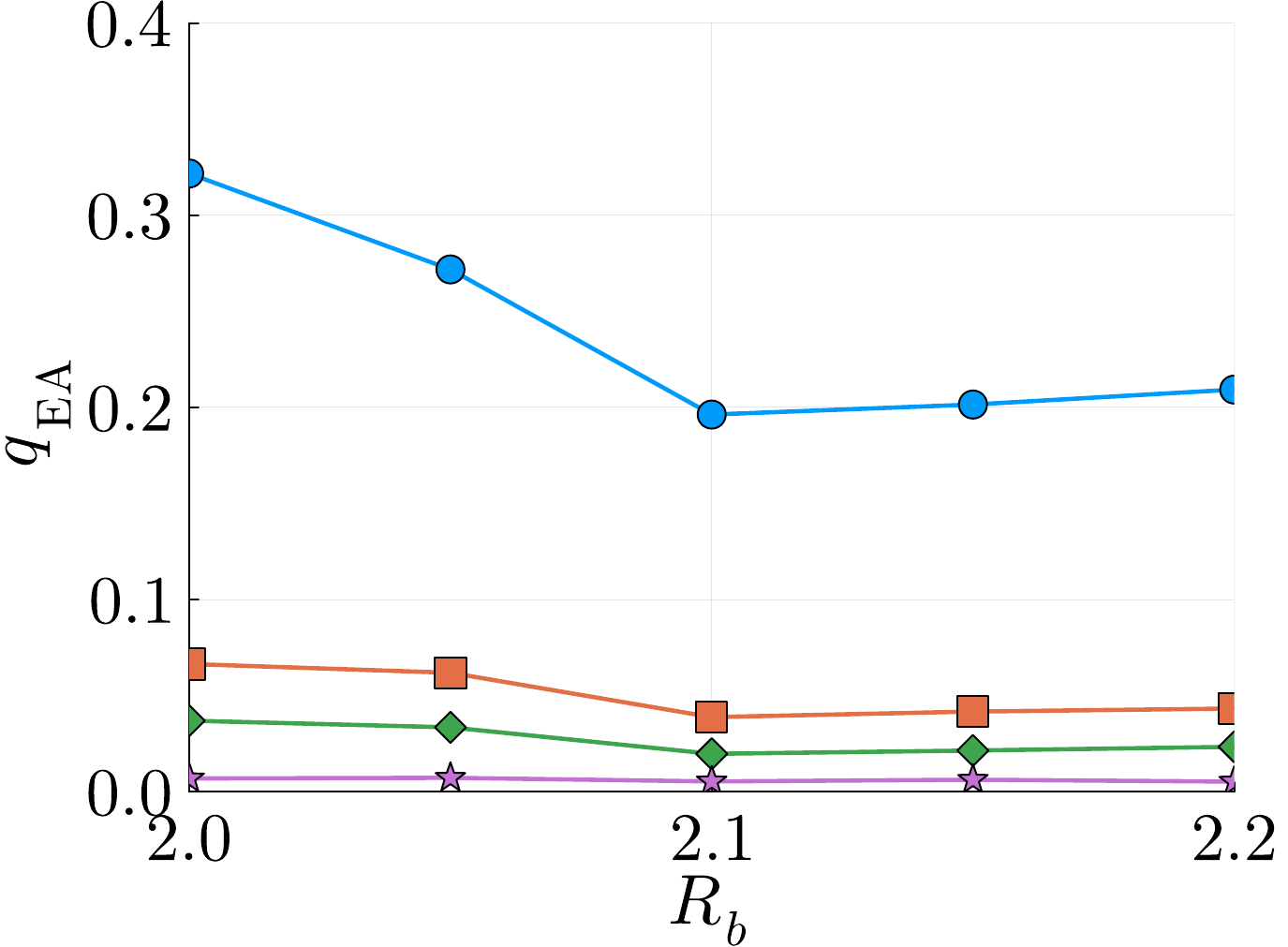}
			\caption{Cut at $\Delta=3.3$}
			\label{fig:EA-cutA}
		\end{subfigure}
		\hfill
		\begin{subfigure}[b]{0.3\textwidth}
			\centering
			\resizebox{\linewidth}{!}{
			\begin{tikzpicture}[x=1.7cm,y=3cm]
				
				\definecolor{leftcol}{RGB}{120,200,200}
				\definecolor{uppercol}{RGB}{220,220,220} 
				\definecolor{lowercol}{RGB}{110,110,110}   
				
				\fill[leftcol]
				(2.4,1.7) rectangle (4.0,2.4);
				
				\fill[lowercol]
				(2.4,1.7)-- (4.0,1.7) -- (4.0,1.97)	-- (3.5,1.97) -- (3.3,1.96)	-- (3.0,1.95) -- (2.5,1.9) -- (2.4,1.88) -- cycle;
				
				\fill[uppercol]
				(2.4,2.3) -- (2.5,2.27) -- (3.0,2.25) -- (3.3,2.25)	-- (3.5,2.25) -- (4.0,2.25)	-- (4.0,2.4) -- (2.4,2.4)
				-- cycle;
				
				\draw[thin]
				(2.4,1.7) rectangle (4.0,2.4);
				
				\draw[thin] (2.4,1.7) -- (4.0,1.7); 
				\draw[thin] (2.4,1.7) -- (2.4,2.4);
				
				\foreach \x in {2.4,2.8,3.2,3.6,4.0} {
					\draw (\x,1.7) -- (\x,1.67);
					\node[below] at (\x,1.67) {\scriptsize $\x$};
				}
				
				\foreach \y in {1.7,1.9,2.1,2.3} {
					\draw (2.4,\y) -- (2.37,\y);
					\node[left] at (2.37,\y) {\scriptsize $\y$};
				}
				
				\draw[dashed, thick]
				plot[smooth] coordinates {(2.4,2.3) (2.5,2.27) (3.0,2.25) (3.3,2.25) (3.5,2.25) (4.0,2.25)};
				
				\draw[dashed, thick]
				plot[smooth] coordinates {(2.4,1.88) (2.5,1.9) (3.0,1.95) (3.3,1.96) (3.5,1.97) (4.0,1.97)};
				
				\node[white] at (3.3,1.81) {\small Nematic};
				\node[white] at (3.3,2.32) {\small Staggered};
				
				\draw[thin,->] (2.4,1.7) -- (4.08,1.7) node[right] {\scriptsize $\Delta$};
				\draw[thin,->] (2.4,1.7) -- (2.4,2.48) node[above] {\scriptsize $R_b$};
				
				\draw[red, thin, ->] (2.5,2.1) -- (4.0,2.1);
				\draw[blue, thin, ->] (3.3,2.0) -- (3.3,2.2);
				
				\node[red, above] at (3.75,2.1) {\scriptsize a)};
				\node[blue, right] at (3.3,2.15) {\scriptsize b)};
			\end{tikzpicture}}
			\caption{Phase diagram}
			\label{fig:Kagome-phasediagram}
		\end{subfigure}
		
		\caption
		{\small	(a) Edwards–Anderson order parameter $q_{\mathrm{EA}}$ as a function of detuning $\Delta$ at fixed $R_b = 2.1$, shown for different boundary-layer cutoffs $b$ that progressively isolate the bulk (see text). The curves correspond to $b=0,1,2,3$, represented by blue circles, red squares, green diamonds, and purple stars, respectively. Successively removing edge sites leads to a rapid suppression of $q_{\mathrm{EA}}$, indicating the absence of glassy order in the bulk. (b) Same quantity as a function of blockade radius $R_b$ at fixed $\Delta = 3.3$, again for varying boundary exclusions, showing consistent decay of $q_{\mathrm{EA}}$ once boundary contributions are excluded. (c) Corresponding section of the phase diagram from \cite{PhysRevLett.130.206501} illustrating the parameter cuts used in (a) and (b).}
		\label{fig:EA}
\end{figure*}

We compute the Edwards-Anderson order parameter \cite{edwards_theory_1975}, which was used in \cite{hibat-allah_recurrent_2025} to distinguish these two possibilities, 
\begin{equation}
 q_\text{EA} = \frac{1}{N} \sum_{i = 1}^{N} \frac{\langle \hat{n}_i - \rho \rangle^2}{\rho (1 - \rho)} 
    \ , \quad \text{where } 
    \rho = \frac{1}{N} \sum_{i = 1}^{N} \langle \hat{n}_i \rangle,
\end{equation}
for different cuts through the intermediate region, as illustrated in Fig.~\ref{fig:Kagome-phasediagram} by the red and blue arrows. 
As discussed above, open boundary conditions favor the formation of pinned excitations near the system edges, leading to finite-size-induced ordering. 
To mitigate this effect, we use the same boundary-layer cutoff $b$ introduced above (Sec. \ref{sec:results_triangular}) and evaluate $q_{\mathrm{EA}}$ only in the remaining bulk region (see Fig.~\ref{fig:EA}). 
While these edge effects could in principle propagate ordering into the bulk, we find that the Edwards-Anderson order parameter rapidly approaches zero once the boundary contributions are removed and remains vanishingly small throughout the entire intermediate region.

Therefore, our findings do not support the existence of a stable glassy phase under open boundary conditions with full long-range interactions. 
Instead, our results support the interpretation of the intermediate regime as a homogeneous disordered phase, in agreement with recent NQS calculations~\cite{hibat-allah_recurrent_2025}, while contrasting with earlier QMC results~\cite{PhysRevLett.130.206501}.

%%%%%%%%%%%%%%%%%%%%%%%%%%%%%%%%%%%%%%%%%%%%%%%%%%%%%%%%%%%%%%%%%%%%%%%%%%%%%%%%%%%%%%%%%%%%%%%%%%%%%%%%
\subsection{Ruby lattice and the topology beyond the PXP approximation}
\label{sec:results_ruby}

Rydberg atom arrays have attracted considerable attention on yet another lattice, namely the Ruby lattice, following the work of Verresen et al. \cite{PhysRevX.11.031005}, who proposed them as a promising platform for realizing a $\mathbb{Z}_2$ topologically ordered phase. 
The Ruby lattice is a Kagomé lattice where the atoms sit on the links instead of the vertices.
A stable topologically ordered ground state for particular values of $\Delta$ has been identified when working in the so-called PXP approximation of the Rydberg potential.
This enforces a hard blockade constraint that forbids simultaneous excitations of sites separated by less than the blockade radius $R_b = 2$, while interactions beyond this distance are neglected. 
In addition to the topologically ordered phase, the phase diagram contains a trivially disordered phase for small $\Delta$ and a valence bond solid (VBS) phase for large $\Delta$, with the topologically ordered phase situated between them.

This topological phase can also be understood from the perspective of quantum dimer models. 
The PXP blockade naturally maps onto a dimer constraint on the underlying Kagomé lattice.
An atom on the edge of the Kagomé lattice can only be excited if no other excited atom shares an adjacent vertex. 
The detuning $\Delta$ controls the density of monomers (vertices not touched by a dimer), and in the limit of large positive $\Delta$ these become energetically suppressed, restricting the low-energy Hilbert space to close-packed dimer coverings. 
Within this manifold, the Rabi term induces resonances between different dimer configurations. 
This results, at 6-th order perturbation theory, in an effective quantum dimer Hamiltonian describing resonances between dimer coverings. 
Since the quantum dimer model on the Kagomé lattice is known to realize a gapped $\mathbb{Z}_2$ topologically ordered ground state \cite{misguich_quantum_2002}, this mapping provides an intuitive explanation for the emergence of topological order in the ruby-lattice PXP model.

Additionally, the low-energy physics of the system admits a theoretical description in terms of a $\mathbb{Z}_2$ lattice gauge theory \cite{samajdar_emergent_2023, sachdev_quantum_2023} with anyonic excitations. 
In this picture, the topological phase corresponds to the deconfined regime, where both types of excitations remain gapped.
Recently, it has also been shown numerically \cite{tarabunga_gauge-theoretic_2022, verresen_unifying_2022} that the ruby-lattice PXP model is adiabatically connected to a stabilizer Hamiltonian consisting of mutually commuting star and plaquette operators that square to one.
This closely parallels Kitaev's toric code model \cite{kitaev_fault-tolerant_2003}, which is the paradigmatic example of a model hosting a $\mathbb{Z}_2$ topologically ordered ground state.

However, the PXP model commits a fairly crude approximation in neglecting interactions beyond the blockade constraint. 
In realistic Rydberg arrays, van-der-Waals interactions remain long-ranged and can compete with the mechanisms stabilizing topological order. 
This possibility was already pointed out by Verresen et al.~\cite{PhysRevX.11.031005}, who found indications from finite-cylinder density matrix renormalization group (DMRG) calculations that longer-range interactions may destabilize the $\mathbb{Z}_2$ spin liquid.
Motivated by this observation, we revisit the question of the stability of the topological phase in the presence of realistic interactions using our sampling-based PEPS framework, which naturally implement the long-ranged interactions and obeys the correct entanglement scaling law. 
In addition to varying the blockade radius $R_b$ and the detuning $\Delta$, we introduce an interaction truncation distance $R_{\mathrm{trunc}}$, which allows us to disentangle the effects of different interaction ranges. 
This provides a systematic way of testing whether the spin-liquid phase survives beyond the PXP limit, as well as whether finite-range approximations can introduce additional spurious phases.

%%%%%%%%%%%%%%%%%%%%%%%%%%%%%%%%%%%%%%%%%%%%%%%%%%%%%%%%%%%%%%%%%%%%%%%%%%%%%%%%%%%%%%%%%%%%%%%%%%%%%%%%
\subsubsection{Detecting topological order}

As topological order cannot be detected by local order parameters, we employ non-local string operators that probe the presence or condensation of emergent $e$ and $m$ anyons in the $\mathbb{Z}_2$ gauge structure (see Appendix~\ref{app:topoEE_string_ops}), following the methodology of Ref.~\cite{PhysRevX.11.031005}. To detect $m$-anyon condensation, we use a ’t Hooft string operator defined along a dual-lattice path $\Gamma^*$,
\begin{equation}
	\hat{T}(\Gamma^*) = \prod_{i \in \Gamma^*} \hat{\sigma}^Z_{i}.
\end{equation}
Open strings create $m$ anyons at their endpoints, so $\hat{T}(\Gamma^*)$ probes phases with condensed $m$ excitations (the VBS phase). Because $\hat{T}(\Gamma^*)$ decays trivially with the length of the path, we use the Fredenhagen–Marcu (FM) order parameter \cite{fredenhagen_charged_1983} to remove this suppression and isolate the contribution from $m$ anyon condensation.
\begin{equation}
	T_{\mathrm{FM}}(\Gamma^*) = \frac{\langle \hat{T}(\Gamma^*) \rangle}{\sqrt{|\langle \hat{T}(\tilde{\Gamma}^*) \rangle|}},
\end{equation}
where $\tilde{\Gamma}^*$ denotes the closed loop obtained by closing the open string $\Gamma^*$ with a path of the same length as the original string operator.

Similarly, $e$-anyon condensation is probed using a Wegner–Wilson string operator defined along a path $\Gamma$ on the direct lattice,
\begin{equation}
	\hat{W}(\Gamma) = \prod_{i \in \Gamma} \hat{\sigma}^X_{i},
\end{equation}
which creates $e$ anyons at its endpoints. The corresponding FM order parameter,
\begin{equation}
	W_{\mathrm{FM}}(\Gamma) = \frac{\langle \hat{W}(\Gamma) \rangle}{\sqrt{|\langle \hat{W}(\tilde{\Gamma}) \rangle|}},
\end{equation}
distinguishes the trivially disordered phase, where $e$ anyons are condensed.

Independently, the topological phase can also be identified through the topological correction \cite{kitaev_topological_2006, levin_detecting_2006} to the area law of entanglement entropy, which is expected to satisfy $\gamma = \log( 2)$ for a $\mathbb{Z}_2$ spin liquid. 
The extraction of the topological entanglement entropy is, however, subject to finite-size corrections and is strictly valid only when the linear size of the bipartitions is much larger than the correlation length. 
Moreover, $\gamma$ is not uniquely fixed within a phase, as additional non-universal contributions can increase its apparent value \cite{kim_universal_2023}. 
These 'spurious' contributions are always non-negative, so $\log(2)$ provides a universal lower bound. This lower bound applies to the asymptotic value of $\gamma$, whereas the finite-size estimates can lie below $\log 2$ due to residual finite-size corrections, which are not constrained to be non-negative.
To evaluate the topological entanglement entropy in our sampling framework, we use the swap trick \cite{PhysRevLett.104.157201}, which computes the second Rényi entropy. 
For non-chiral topological order, the topological correction $\gamma$ is independent of the Rényi index \cite{flammia_topological_2009}, so that all Rényi entropies yield the same value for the topological correction. 
In practice, $\gamma$ is extracted from linear combinations of subsystem entropies following the Kitaev–Preskill construction, while string-operator expectation values used to diagnose anyon condensation are computed as described in Appendix~\ref{app:topoEE_string_ops}. 
For the accessible subsystem sizes considered here, the Kitaev–Preskill~\cite{kitaev_topological_2006} construction is preferred over the Levin–Wen scheme~\cite{levin_detecting_2006}, as the boundaries of the subsystems are more widely separated, thereby reducing finite-size effects. 
In contrast, the ring-shaped partitions in the Levin–Wen construction involve subsystem sizes comparable to the correlation length, which enhances finite-size errors.

%%%%%%%%%%%%%%%%%%%%%%%%%%%%%%%%%%%%%%%%%%%%%%%%%%%%%%%%%%%%%%%%%%%%%%%%%%%%%%%%%%%%%%%%%%%%%%%%%%%%%%%%
\subsubsection{Results within the PXP approximation}

\begin{figure}
    \centering
    \includegraphics[width=0.99\linewidth]{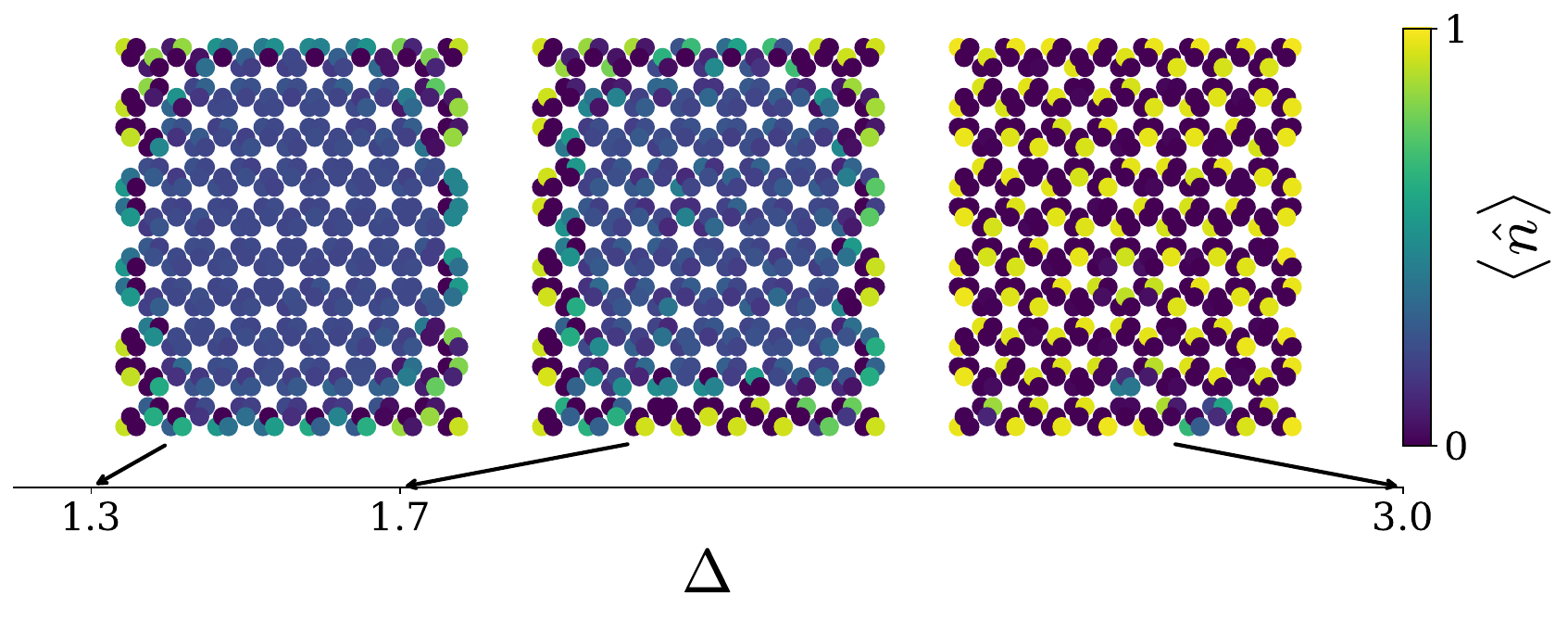}
    \caption{\small Average local excitation density per site for three representative phases of the PXP model phase diagram. As expected, this local observable does not distinguish between the trivially disordered ($\Delta = 1.3$) and the topologically ordered ($\Delta = 1.7$) phases.}
    \label{fig:PXP_excitationdensities}
\end{figure}

\begin{table}
	\centering
	\renewcommand{\arraystretch}{1.4}
	\begin{tabular}{|c|c|c|c|}
		\hline
	\textbf{Phase} & $W_\text{FM}$ & $T_\text{FM}$ & $\gamma_{\mathrm{KP}} / \gamma_{\mathbb{Z}_2}$ \\ %\text{Kitaev-Preskill}
	\hline
	Trivial & $0.22 \pm 0.01$ & $0.06 \pm 0.04$ & $0.12 \pm 0.09$ \\ \hline
 %$0.08 \pm 0.06$\\ \hline
	Topological & $0.022 \pm 0.012$ & $0.026 \pm 0.01$ & $0.91 \pm 0.12$\\ \hline
 %$0.63 \pm 0.08$\\ \hline
	VBS & $0.0002 \pm 0.0001$ & $1.008 \pm 0.006$ & $-0.006 \pm 0.012$\\ \hline
 %$-0.004 \pm 0.008$\\ \hline
	\end{tabular}
	\caption{Fredenhagen–Marcu order parameters and finite-size estimates of the topological entanglement entropy $\gamma_{\mathrm{KP}}$, obtained from the Kitaev–Preskill construction and expressed in units of $\gamma_{\mathbb{Z}_2}\equiv\log 2$, for representative states in different phases of the PXP model (see Fig.~\ref{fig:PXP_excitationdensities}).}
	\label{tab:PXP_FMOP}
\end{table}

Before investigating the role of realistic interactions, we first benchmark our PEPS simulations against the PXP phase diagram obtained by Verresen et al. \cite{PhysRevX.11.031005}. Our results (Fig.~\ref{fig:PXP_excitationdensities}, Tab.~\ref{tab:PXP_FMOP}) reproduce the previously reported phase structure, including a $\mathbb{Z}_2$ topologically ordered phase at intermediate detuning $\Delta$, as well as adjacent VBS and trivially disordered phases. 
The presence of topological order is confirmed by vanishing Fredenhagen–Marcu order parameters in the intermediate regime, together with a finite topological entanglement entropy.
These diagnostics are consistent with a $\mathbb{Z}_2$ spin liquid and validate our numerical approach in the constrained limit. 
Importantly, the other two phases show no sign of topological ordering, indicated by the vanishing topological entanglement entropy and one non-zero Fredenhagen-Marcu order parameter. 
It is particularly noteworthy that the topologically ordered phase remains stable even in the presence of open boundary conditions. 
Although the boundaries of the finite system are partially pinned into a crystalline configuration, topological order persists in the bulk.

The emergence of the $\mathbb{Z}_2$ topologically ordered phase in the PXP limit can also be understood from the tensor-network perspective.
In Appendix~\ref{app:one_parameter_ansatz}, we construct a variational PEPS ansatz motivated by the exact representability of the toric code wavefunction as a tensor network with bond dimension two \cite{zeng_quantum_2018}. 
The ansatz is built from local tensors enforcing the PXP constraint, while a single variational parameter $p$ captures fluctuations beyond the ideal toric-code limit. 
The physical motivation for including such fluctuations follows from the fact that the presence of monomer configurations allows a single-site kinetic term (the Rabi term) to generate the multisite resonances between dimer configurations required to stabilize the spin-liquid phase \cite{PhysRevX.11.031005}.
Our ansatz provides a simple analytical framework for incorporating these fluctuations to understand and study the topological phase.

%%%%%%%%%%%%%%%%%%%%%%%%%%%%%%%%%%%%%%%%%%%%%%%%%%%%%%%%%%%%%%%%%%%%%%%%%%%%%%%%%%%%%%%%%%%%%%%%%%%%%%%%
\subsubsection{Realistic Rydberg interactions}

\begin{figure}[b]
    \centering
    \includegraphics[width=0.9\linewidth]{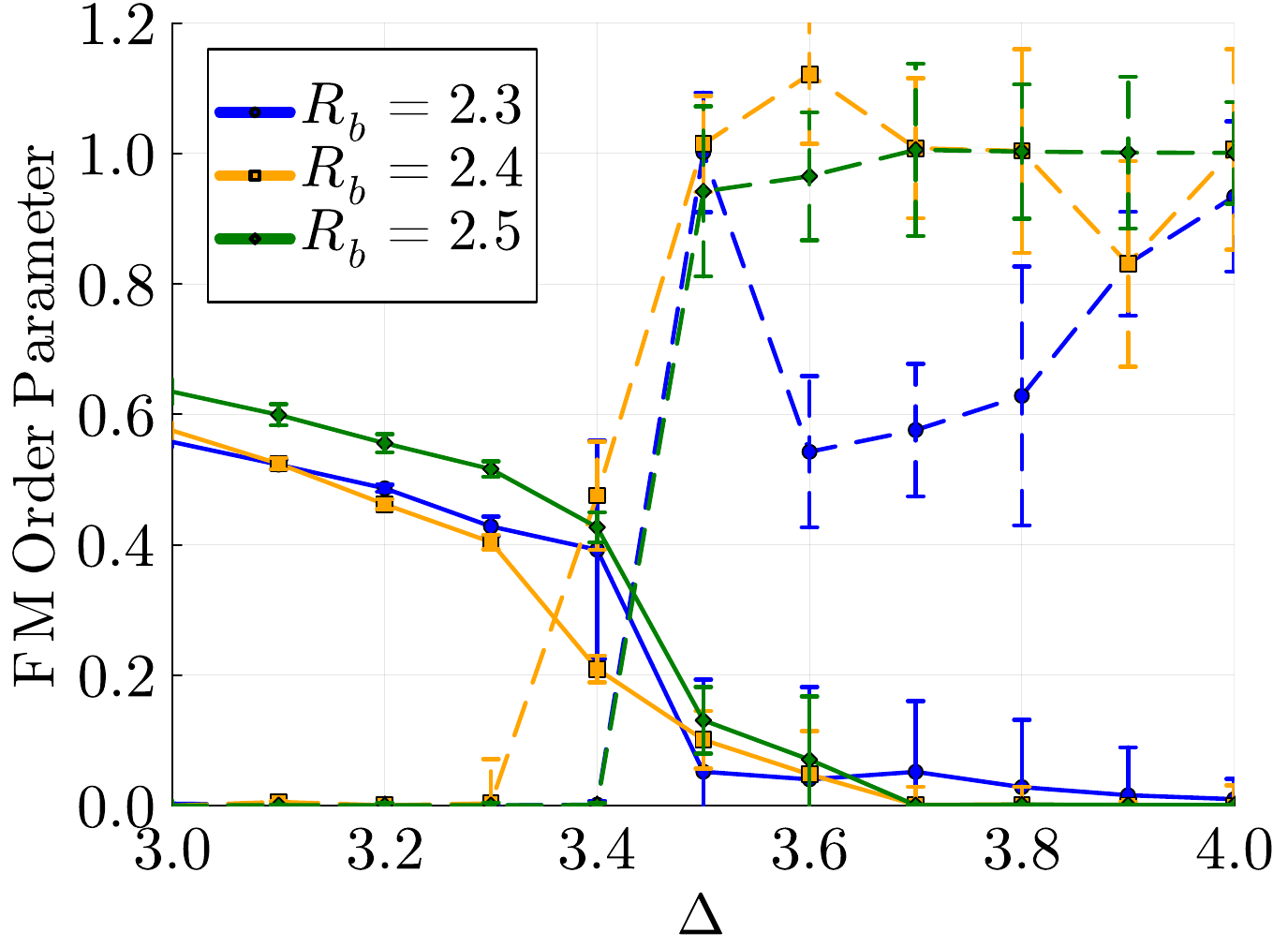}
    \caption{\small 
    Fredenhagen–Marcu order parameters $W_{\mathrm{FM}}$ (Wegner–Wilson loop, solid lines) and $T_{\mathrm{FM}}$ (’t Hooft loop, dashed lines) as a function of detuning $\Delta$ for three values of the blockade radius $R_b \in \{2.3, 2.4, 2.5\}$, in the absence of any truncation of the interactions in the numerical treatment. 
    The different line colors/markers correspond to increasing $R_b$ (blue/circle: $R_b=2.3$, orange/square: $R_b=2.4$, green/diamond: $R_b=2.5$). 
    For all cases, the system shows a direct crossover from an $e$-anyon condensate (trivial phase) at low detuning, indicated by large $W_{\mathrm{FM}}$ and small $T_{\mathrm{FM}}$, to an $m$-anyon condensate (VBS phase) at larger $\Delta$, indicated by small $W_{\mathrm{FM}}$ and large $T_{\mathrm{FM}}$, without an intermediate regime where both order parameters are simultaneously suppressed.  
    This indicates the absence of an intervening topologically ordered phase in the presence of realistic long-range interactions.
    } \label{fig:FMOP-fullinter}
\end{figure}

Here we turn to the systematic treatment of realistic interactions and the effects of the truncation radius in their numerical treatment.
We start by not truncating interactions at all, a feature that our PEPS algorithm allows for.
Across the investigated range $R_b \in [2.3, 2.5]$, the Fredenhagen–Marcu order parameters (Fig.~\ref{fig:FMOP-fullinter}) show a direct crossover from an $e$-condensed trivial phase at low detuning to an $m$-condensed VBS phase at larger $\Delta$, with no intermediate regime where both $T_{\mathrm{FM}}$ and $W_{\mathrm{FM}}$ are simultaneously small.
This indicates the absence of a stable topologically ordered phase in the presence of realistic interactions within the investigated parameter region.
Additionally, the topological entanglement entropy obtained from the Kitaev-Preskill construction shows no finite topological contribution, with all measured values in the range of $\gamma \in \left[-0.1, 0.04\right]$ consistent with zero within the estimated error bounds.

The different values for the $T_\mathrm{FM}$ order parameter for $R_b = 2.3$ and $\Delta \in [3.5, 4.0]$ can be attributed to different boundary crystallizations, which lead to a different ordering in the bulk. 
When initializing all points with the same excitations pinned at the boundary, the optimized wavefunctions show very similar values of $T_\mathrm{FM}$.

It is important to note that the interpretation of the Fredenhagen-Marcu order parameter relies on the validity of the underlying gauge constraint, which is exact only in the strong blockade limit. 
For sufficiently small $R_b$, additional constraint-violating configurations (excitations separated by exactly two lattice spacings) become energetically accessible and the mapping onto the effective gauge theory breaks down. 
Consequently, the Fredenhagen-Marcu order parameters no longer provide a reliable distinction between the three phases. 
This starts happening for $R_b \leq 2.2$, implying that this regime cannot be reliably used to diagnose the presence of topological order using the effective gauge-theory description. 
On the other hand, increasing the blockade radius to $R_b \geq 2.6$ enhances geometric constraints and stabilizes crystalline order, leaving little room for the fluctuations required to sustain a topologically ordered spin liquid.
Both limits therefore disfavour the deconfined phase, consistent with related tendencies already observed in the cylinder DMRG calculations of Verresen et al. \cite{PhysRevX.11.031005}.

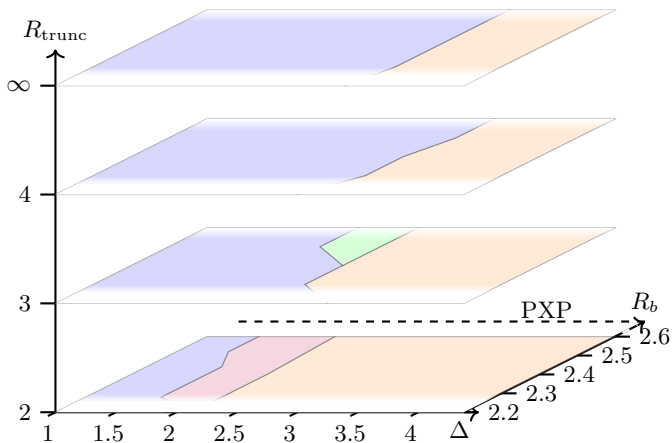
\begin{figure}[htbp]
        \vspace{4mm}
		\centering
		\begin{tikzpicture}[scale=4,x={(1cm,0cm)},y={(0cm,1.5cm)},z={(-0.5cm,-0.25cm)}]
			\draw[->,thick] ({-0.5/2.5},0.2,0) -- (1.2,0.2,0) node[anchor=north east] {$\Delta$}; 
			\draw[->,thick] (1.15,0.2,0) -- (1.15,0.2,-1.2) node[anchor=south] {$R_b$};  
			\draw[->,thick] ({-0.5/2.5},0.2,0) -- ({-0.5/2.5},1,0) node[above] {$R_\mathrm{trunc}$};
			
			\draw[dashed,thick, white] (0,0.8,0) -- (0,0.9,0);
			\draw[dashed,thick, white] (1.15,0.2,-1.05) -- (1.15,0.2,-1.15);
			
			\foreach \x in {-0.2,0,0.2,0.4,0.6,0.8,1}{
				\draw[thick] (\x, 0.2, 0) -- (\x, 0.2, 0.05) node[below] {$\pgfmathprintnumber{\fpeval{1.5 + 2.5*\x}}$};
			}
			
			\foreach \x in {0, 0.25, 0.5, 0.75, 1}{
				\draw[thick] (1.15, 0.2, -\x) -- (1.2, 0.2, -\x) node[right] {$\pgfmathprintnumber{\fpeval{{2.2 + \x*4/10}}}$};
			}
			
			\draw[thick] ({-0.5/2.5}, 0.2, 0) -- ({-0.5/2.5 -0.05}, 0.2, 0) node[left] {$2$};
			\draw[thick] ({-0.5/2.5}, 0.44, 0) -- ({-0.5/2.5 -0.05}, 0.44, 0) node[left] {$3$};
			\draw[thick] ({-0.5/2.5}, 0.68, 0) -- ({-0.5/2.5 -0.05}, 0.68, 0) node[left] {$4$};
			\draw[thick] ({-0.5/2.5}, 0.92, 0) -- ({-0.5/2.5 -0.05}, 0.92, 0) node[left] {$\infty$};

			\filldraw[fill=blue!35,draw=black!70,opacity=0.45]
			({-0.5/2.5},0.2,0) -- ({0.375/2.5},0.2,0) -- ({0.375/2.5},0.2,-0.1) -- ({0.125/2.5},0.2,-0.2) -- ({0.125/2.5},0.2,-0.6) -- ({-0.07/2.5},0.2,-0.8) -- ({-0.07/2.5},0.2,-1) -- ({-0.5/2.5},0.2,-1) -- cycle;
			\filldraw[fill=purple!35,draw=black!70,opacity=0.45]
			({0.375/2.5},0.2,0) -- ({0.375/2.5},0.2,-0.1) -- ({0.125/2.5},0.2,-0.2) -- ({0.125/2.5},0.2,-0.6) -- ({-0.07/2.5},0.2,-0.8) -- ({-0.07/2.5},0.2,-1) -- ({0.56/2.5},0.2,-1) -- ({0.625/2.5},0.2,-0.5) -- ({0.625/2.5},0.2,-0.25) -- ({0.625/2.5},0.2,-0.1) -- cycle;
			\filldraw[fill=orange!35,draw=black!70,opacity=0.45]
			({0.56/2.5},0.2,-1) -- ({0.625/2.5},0.2,-0.5) -- ({0.625/2.5},0.2,-0.25) -- ({0.625/2.5},0.2,-0.1) -- ({0.625/2.5},0.2,-0.1) -- ({0.5/2.5},0.2,0) -- (1.15, 0.2, 0) -- (1.15, 0.2, -1) -- cycle;

			\filldraw[fill=blue!35,draw=black!70,opacity=0.45]
			({-0.5/2.5},0.68,0) -- ({1.5/2.5},0.68,0) -- ({1.75/2.5},0.68,-0.25) -- ({1.75/2.5},0.68,-0.5) -- ({1.875/2.5},0.68,-0.75) -- ({1.875/2.5},0.68,-1)  -- ({-0.5/2.5},0.68,-1) -- cycle;
			\filldraw[fill=orange!35,draw=black!70,opacity=0.45]
			(1.15,0.68,0) -- ({1.5/2.5},0.68,0) -- ({1.75/2.5},0.68,-0.25) -- ({1.75/2.5},0.68,-0.5) -- ({1.875/2.5},0.68,-0.75) -- ({1.875/2.5},0.68,-1)  -- (1.15,0.68,-1) -- cycle;
			
			\filldraw[fill=blue!35,draw=black!70,opacity=0.45]
			({-0.5/2.5},0.44,0) -- ({1.75/2.5},0.44,0) -- ({1.25/2.5},0.44,-0.25) -- ({1.25/2.5},0.44,-0.5) -- ({0.75/2.5},0.44,-0.75) -- ({0.75/2.5},0.44,-1) -- ({-0.5/2.5},0.44,-1) -- cycle; 
			
			\filldraw[fill=orange!35,draw=black!70,opacity=0.45]
			(1.15,0.44,0) -- ({1.75/2.5},0.44,0) -- ({1.25/2.5},0.44,-0.25) -- ({1.25/2.5},0.44,-0.5) -- ({1.25/2.5},0.44,-0.75) -- ({1.25/2.5},0.44,-1)  -- (1.15,0.44,-1) -- cycle; 
			
			\filldraw[fill=green!35,draw=black!70,opacity=0.45]
			({1.25/2.5},0.44,-0.5) -- ({0.75/2.5},0.44,-0.75) -- ({0.75/2.5},0.44,-1) -- ({1.25/2.5},0.44,-1) -- ({1.25/2.5},0.44,-0.75) -- cycle;

			\filldraw[fill=blue!35,draw=black!70,opacity=0.45]
			({-0.5/2.5},0.92,0) -- ({1.9/2.5},0.92,0) -- ({2.0/2.5},0.92,-0.25) -- ({2.0/2.5},0.92,-0.5) -- ({2.0/2.5},0.92,-0.75) -- ({2.0/2.5},0.92,-1)  -- ({-0.5/2.5},0.92,-1) -- cycle;
			\filldraw[fill=orange!35,draw=black!70,opacity=0.45]
			(1.15,0.92,0) -- ({1.9/2.5},0.92,0) -- ({2.0/2.5},0.92,-0.25) -- ({2.0/2.5},0.92,-0.5) -- ({2.0/2.5},0.92,-0.75) -- ({2.0/2.5},0.92,-1)  -- (1.15,0.92,-1) -- cycle;
			
			\tikzfading[name=fadeup,
			top color=transparent!100,
			bottom color=transparent!0]
			\tikzfading[name=fadedown,
			top color=transparent!0,
			bottom color=transparent!100]
			\def\fadelim{-0.15}
			\foreach \y in {0.2, 0.44, 0.68, 0.92}{
				\fill[white,path fading=fadeup]
				({-0.5/2.5},\y,\fadelim) -- (1.15,\y,\fadelim) -- ({1.15},\y,-0.25) -- ({-0.5/2.5},\y,-0.25) -- cycle;
				\fill[white]
				({-0.5/2.5},\y,\fadelim) -- (1.15,\y,\fadelim) -- ({1.15},\y,0) -- ({-0.5/2.5},\y,0) -- cycle;
			}
			
			\foreach \y in {0.44, 0.68, 0.92}{
				\fill[white,path fading=fadedown]
				({1.15},\y,-1) -- ({-0.5/2.5},\y,-1) -- ({-0.5/2.5},\y,-\fadelim-1) -- (1.15,\y,-\fadelim-1) --  cycle;
		}
			
			\draw[dashed,thick] (1.15,0.2,-1.2) -- ({-0.5/2.5},0.2,-1.2);
			\node at (0.75,0.2,-1.35) {PXP};
			
		\end{tikzpicture}
		\caption
		{\small Three-dimensional phase diagram showing the phase structure as a function of $R_{\mathrm{trunc}}$, $\Delta$, and $R_b$. The trivial, VBS, topological, and additional $R_{\mathrm{trunc}}=3$ phases are shown in blue, orange, purple, and green, respectively. For small $R_b$ (white), constraint-violating configurations become relevant, causing the effective gauge-theory description and the corresponding Fredenhagen-Marcu order parameter classification to break down. The phase boundaries are drawn halfway between neighboring data points assigned to different phases. The underlying data points are omitted for clarity.} 
		\label{fig:ThreeDimPD}
	\end{figure}

We next investigate how this result depends on the interaction range by introducing a truncation distance $R_{\mathrm{trunc}}$. Realistic van der Waals interactions are retained up to distance $R_{\mathrm{trunc}}$ and set to zero beyond this range. The PXP model corresponds to the limit $R_{\mathrm{trunc}}=2$ and $R_b\rightarrow\infty$, while the fully interacting model considered above is recovered for $R_{\mathrm{trunc}}\rightarrow\infty$ at finite $R_b$. We consider $R_{\mathrm{trunc}} \in \{2,3,4,\infty\}$, as shown in Fig.~\ref{fig:ThreeDimPD}.

For $R_{\mathrm{trunc}}=2$, we find a topologically ordered ground state even away from the PXP limit, indicating that a strong but finite repulsion is enough to stabilize the $\mathbb{Z}_2$ liquid when truncating longer-ranged interactions. 
However, this phase does not remain stable when including additional interaction terms, as all other truncation schemes show no signatures of topological order, similarly to what is shown above for $R_{\mathrm{trunc}}=\infty$. 
Thus, we confirm that longer-range interactions destabilize the spin-liquid phase.
Our results provide an explicit demonstration of this effect by systematically comparing different interaction truncations.

\begin{figure}[t]
    \centering
		\begin{subfigure}[b]{0.23\textwidth}
			\centering
			\includegraphics[width=1.1\textwidth]{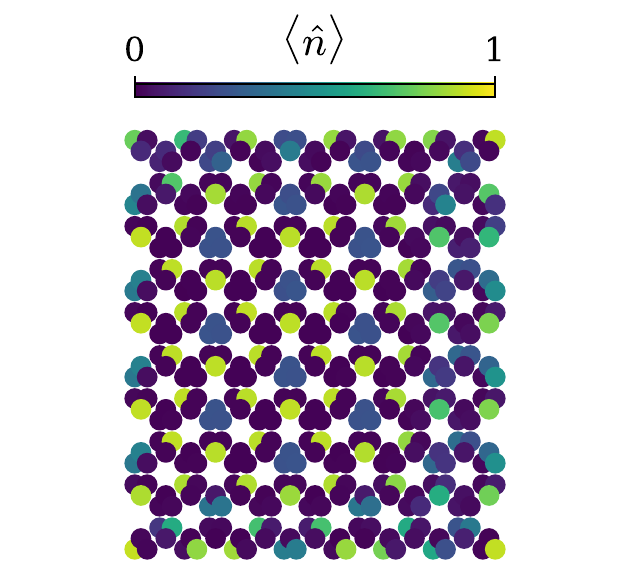}			
	
		\end{subfigure}
		\hfill
		\begin{subfigure}[b]{0.23\textwidth}
			\centering
			\includegraphics[width=\textwidth]{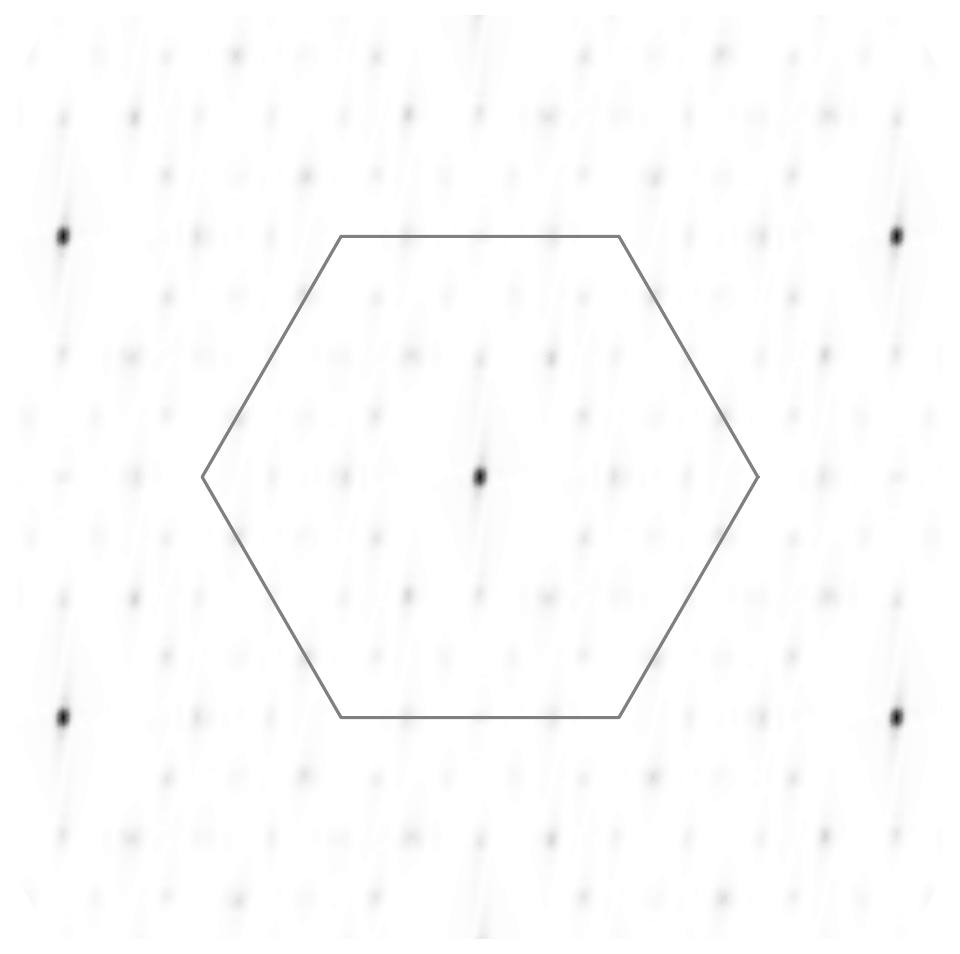}
			
		\end{subfigure}
		\caption{\small Local excitation density (left) and structure factor (right) of the additional phase highlighted in green in Fig.~\ref{fig:ThreeDimPD}. The phase is characterized by homogeneous regions embedded in a background of pinned crystalline excitations. The hexagon indicates the fundamental region in reciprocal space containing the non-redundant information of the structure factor. Owing to the six-atom basis of the ruby lattice, this region does not coincide with the first Brillouin zone of the underlying triangular Bravais lattice.}
        \label{fig:Rb25_Delta25_Rtrunc3}
 \end{figure}

Furthermore, within the ($R_{\mathrm{trunc}}=3$) regime, we identify an additional phase whose excitation density is shown in Fig.~\ref{fig:Rb25_Delta25_Rtrunc3}. 
This phase is particularly interesting, as it exhibits pinned crystalline excitations coexisting with completely homogeneous regions. To compute the structure factor (Fig.~\ref{fig:Rb25_Delta25_Rtrunc3}, we treat the Ruby lattice as a triangular lattice with a 6-atom basis (cf. \cite{maity_fermionic_2026}).

Such a coexistence of spatially ordered and disordered features is not observed for the other truncation schemes. 
The fact that this phase disappears when modifying the interaction range further emphasizes the potential biases introduced by truncated interactions on a specific lattice and demonstrates that finite-range approximations can generate artificial phases that are absent in the full interaction model.
\section{Conclusion and Outlook}

In this work, we have showcased the possibility of combining the native two-dimensional entanglement structure of finite PEPS with realistic lattice geometries (including boundaries) and full long-range interactions.
We have shown that these details can qualitatively affect the phase diagram of Rydberg atom arrays on different lattice structures.

In particular, we find no evidence for a stable spin-glass phase on the Kagomé lattice, and we demonstrate that the $\mathbb{Z}_2$ topologically ordered phase of the Ruby-lattice PXP model is destabilized once realistic long-range interactions are included.
Both results confirm previous conjectures in the literature and highlight the sensitivity of these exotic phases to realistic microscopic details, underscoring the challenges involved in realizing them in experimentally relevant settings.

Our results highlight the importance of studying Rydberg models beyond idealized effective descriptions. 
While topological order can emerge in constrained Rydberg models, realizing such phases as robust ground states of experimentally realistic systems remains a challenge. This motivates the exploration of extended Rydberg platforms and alternative interaction schemes that may provide additional mechanisms for stabilizing topological phases. 
One such route may be multispecies RAAs \cite{beterov_rydberg_2015, anand_dual-species_2024, cesa_engineering_2026}, which could enable interaction schemes that more closely mimic the PXP limit and thereby allow topological phases to emerge.

Moreover, the specific features of PEPS with a sampling-based optimization make them appealing as a complementary route for constructing reliable digital twins of quantum simulation experiments. We could envision the possibility of expanding the capabilities of already existing remarkable examples based on Tree Tensor Networks~\cite{felser_efficient_2021, jaschke_ab-initio_2024}.

\section*{Acknowledgments}

The simulations presented in this work were produced with a code based on \cite{schroder_netenesyqumaquantumnaturalfpepsjl_2025, puente_netenesyqumaquantumnaturalgradientjl_2025}. Parts of them make use of the ITensor library \cite{itensor, itensor-r0.3}. 
Data and code will be made publicly available on Zenodo upon completion of the peer-review process.
The authors acknowledge the support from the DFG under Germany's Excellence Strategy - Cluster of Excellence Matter and Light for Quantum Computing (ML4Q) EXC 2004\slash 2 – 390534769, and project Grant No. 277101999 within the CRC network TR 183.
Support by the EU via the Horizon Europe programme HORIZON-CL4-
2022-QUANTUM-02-SGA via the project 101113690
(PASQuanS2.1) is also acknowledged.
The authors gratefully acknowledge the Gauss Centre for Supercomputing e.V. (GCS, \href{www.gauss-centre.eu}{www.gauss-centre.eu}) for providing computing time through the John von Neumann Institute for Computing (NIC) on the GCS Supercomputer JUPITER, and the FZJ/VSR for computing time on JURECA~\cite{JURECA2021} at Jülich Supercomputing Centre (JSC) under grant no. SyQMa2TeNs.
JUPITER is supported by the EuroHPC JU and GCS through funding by the European Commission, the German Federal Ministry of Research, Technology and Space, and the Ministry of Culture and Science of the State of North Rhine-Westphalia.

\bibliography{literature.bib}
\clearpage

\appendix
\section{Mapping from triangular to square lattice} \label{app:triangularmapping}

\begin{figure}[t]
\centering
\resizebox{\columnwidth}{!}{
\begin{tikzpicture}[scale=0.55, every node/.style={draw, fill, circle, inner sep=1pt}]

\begin{scope}
\def\Lval{1}
\pgfmathsetmacro{\LminusOne}{\Lval-1}

\foreach \x in {-\Lval,...,\Lval}{
    \draw[thick]
    ({\x - 1/2*\Lval},{-sqrt(3)/2*\Lval}) --
    ({\x + 1/2*\Lval},{sqrt(3)/2*\Lval});

    \draw[thick]
    ({\x/2 - \Lval},{\x*sqrt(3)/2}) --
    ({\Lval + \x/2},{\x*sqrt(3)/2});
}

\draw[thick] ({1},{sqrt(3)/2*0}) -- ({0+1/2},{sqrt(3)/2*1});
\draw[thick] ({0},{sqrt(3)/2*0}) -- ({-1+1/2},{sqrt(3)/2*1});
\draw[thick] ({1/2},{sqrt(3)/2*-1}) -- ({-1/2+1/2},{sqrt(3)/2*0});
\draw[thick] ({-1/2},{sqrt(3)/2*-1}) -- ({-1},{sqrt(3)/2*0});

\foreach \x in {-\Lval,...,\Lval}{
    \foreach \y in {-\Lval,...,\Lval}{
        \node[fill=blue!50, minimum size=0.18cm]
        at ({\x+\y/2},{sqrt(3)/2*\y}) {};
    }
}

\end{scope}

\begin{scope}
\draw[->] (2,0) -- (3,0);
\end{scope}

\begin{scope}[xshift=5cm]

\def\Lval{1}
\pgfmathsetmacro{\LminusOne}{\Lval-1}

\foreach \x in {-\Lval,...,\Lval}{
    \foreach \y in {-\Lval,...,\Lval}{
        \node (T\x\y) at (\x,\y) {};
        \node[fill=blue!50, minimum size=0.18cm] at (\x,\y) {};
    }
}

\foreach \x in {-\Lval,...,\LminusOne}{
    \foreach \y in {-\Lval,...,\Lval}{
        \pgfmathtruncatemacro{\nx}{\x+1}
        \draw[thick] (T\x\y) -- (T\nx\y);
    }
}

\foreach \x in {-\Lval,...,\Lval}{
    \foreach \y in {-\Lval,...,\LminusOne}{
        \pgfmathtruncatemacro{\ny}{\y+1}
        \draw[thick] (T\x\y) -- (T\x\ny);
    }
}

\foreach \x in {-\Lval,...,\LminusOne}{
    \foreach \y in {-\Lval,...,\LminusOne}{
        \pgfmathtruncatemacro{\nx}{\x+1}
        \pgfmathtruncatemacro{\ny}{\y+1}
        \draw[thick, dashed] (T\x\ny) -- (T\nx\y);
    }
}

\foreach \x in {-\Lval,...,\Lval}{
    \foreach \y in {-\Lval,...,\Lval}{
        \node (T\x\y) at (\x,\y) {};
        \node[fill=blue!50, minimum size=0.18cm] at (\x,\y) {};
    }
}

\end{scope}

\end{tikzpicture}
}
\caption{Mapping of the triangular lattice onto the square-lattice PEPS geometry. The reduced coordination number implies that some bonds cannot be represented faithfully (dashed).}
\label{fig:triangulartosquare}
\end{figure}
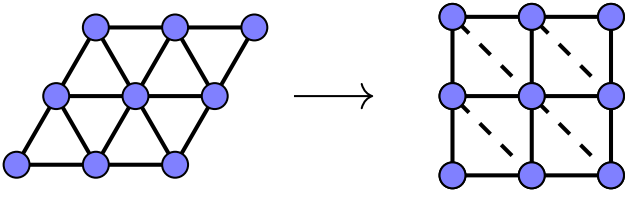

In order to apply our algorithm, we employ a one-to-one mapping of the triangular lattice sites onto the square lattice sites shown in Fig.~\ref{fig:triangulartosquare}.
This embedding introduces some approximations: sites that are nearest neighbours on the triangular lattice can be mapped to next-nearest neighbours on the square lattice, such that their correlations must be mediated indirectly through intermediate sites. 
Consequently, vertical and horizontal bonds are treated differently from diagonal ones, which breaks the 6-fold rotational symmetry of the original lattice. 
A sufficiently large bond dimension should nevertheless allow the variational optimization to capture the correct physics. 
The clear six-fold rotational symmetry of the structure factors in \ref{fig:sf-triangular} already indicates that the physics is correctly described. To further verify that the mapping does not introduce a significant bias, we compare the mutual information $I_{AB}$ defined by,
\begin{equation}
	I_{AB} = S_A + S_B - S_{AB},
\end{equation}
between the two diagonal pairs of sites in a square-lattice plaquette in Table~\ref{tab:mutualinformation}. 

\begin{table}[h!]
    \centering
     \begin{tikzpicture}[
    scale=0.75,
    site/.style={draw, fill=blue!50, circle,
                 minimum size=0.18cm, inner sep=1pt},
    lab/.style={font=\scriptsize, inner sep=0pt}
]
    \node[site] (A) at (0,1) {$A$};
    \node[site] (B) at (1,1) {$B$};
    \node[site] (C) at (0,0) {$C$};
    \node[site] (D) at (1,0) {$D$};

    % Square-lattice bonds
    \draw[thick] (A)--(B)--(D)--(C)--(A);

    % Pairs of interest
    \draw[thick,dashed] (A)--(D);
\end{tikzpicture}

    \vspace{2pt}
    \caption{Mutual information for physically connected (disconnected) diagonal pairs $AD$ ($BC$) across different parameter sets.}
    \label{tab:mutualinformation}
    \renewcommand{\arraystretch}{1.4}
    \begin{tabular}{|c|c|c||c|c|}
        \hline
        \multicolumn{3}{|c||}{\textbf{Phase Parameters}}
        & \multicolumn{2}{c|}{\textbf{Mutual Information}} \\ \hline
        \textbf{Name} & $\Delta$ & $R_b$
        & $I_{AD}\,[10^{-3}]$ & $I_{BC}\,[10^{-3}]$ \\ \hline
        $\frac{1}{3}$-filling & 2.0 & 1.2
        & $2.502 \pm 0.277$ & $0.241 \pm 0.038$ \\ \hline
        $\frac{1}{4}$-filling & 3.0 & 1.6
        & $0.011 \pm 0.003$ & $0.005 \pm 0.001$ \\ \hline
        Disordered & 2.0 & 1.8
        & $40.389 \pm 4.01$ & $4.361 \pm 0.6$ \\ \hline
        Disordered & 2.5 & 1.0
        & $32.65 \pm 0.706$ & $8.134 \pm 0.284$ \\ \hline
        $\frac{1}{7}$-filling & 3.5 & 2.1
        & $1.631 \pm 0.777$ & $0.094 \pm 0.06$ \\ \hline
    \end{tabular}
\end{table}

Although neither pair is connected by a virtual bond, the pair that corresponds to nearest neighbours on the original triangular lattice exhibits significantly stronger correlations than the physically disconnected pair. 
Thus, the optimization recovers the relevant correlations of the triangular lattice despite the reduced connectivity.

%%%%%%%%%%%%%%%%%%%%%%%%%%%%%%%%%%%%%%%%%%%%%%%%%%%%%%%%%%%%
\begin{figure}[htbp]
    \centering
    \includegraphics[width=0.9\linewidth]{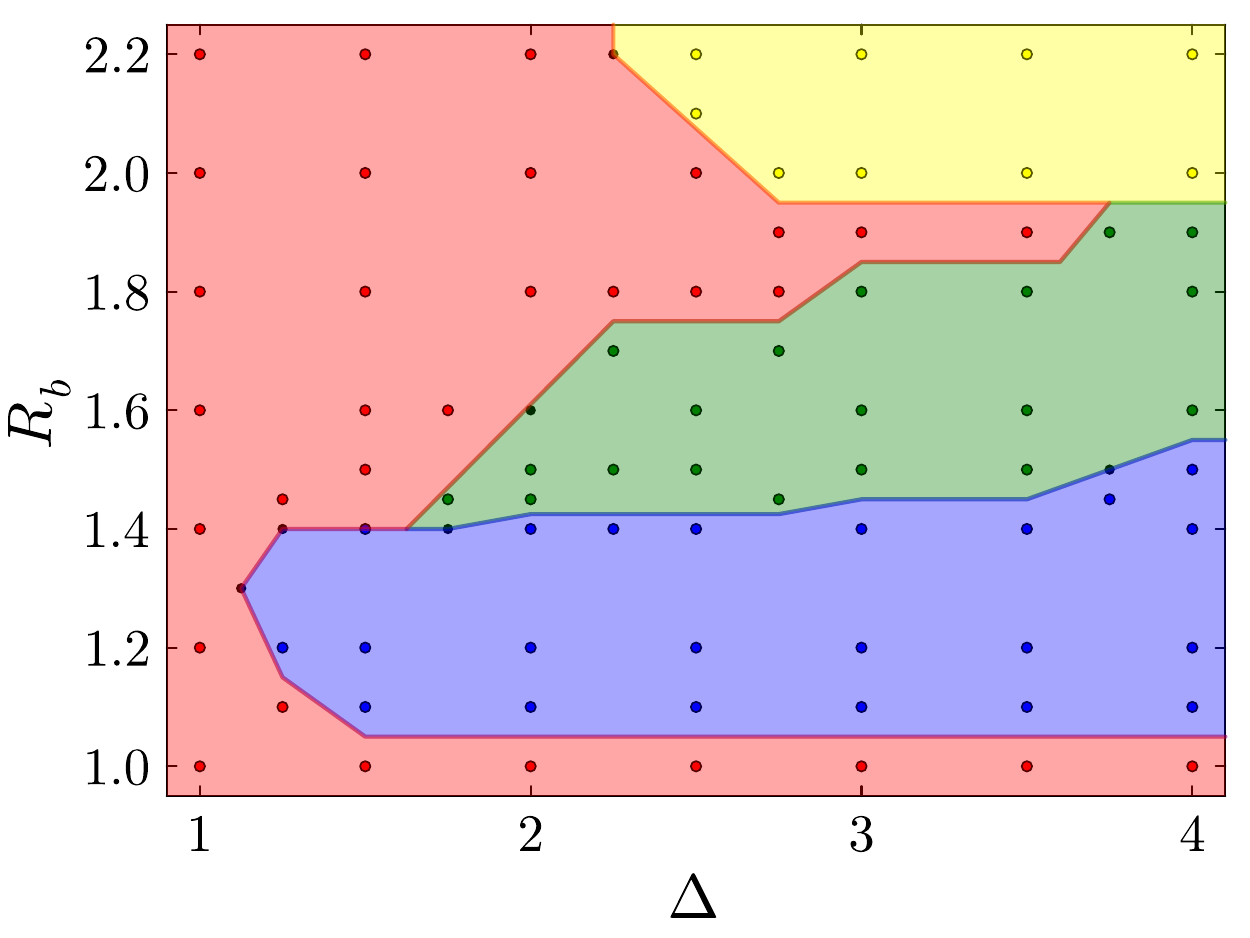}
    \caption{Sketchy phase diagram of the Rydberg atom array model on the triangular lattice with open boundary conditions. The dots indicate simulated phase space points. In addition to the quantum disordered phase with homogeneous bulk density (\textit{red}), we find several density wave phases. At small values of $R_b$, we identify a $\frac{1}{3}$-filling density wave (\textit{blue}), while at higher values we further identify a density wave phase at $\frac{1}{4}$-filling (\textit{green}) and $\frac{1}{7}$-filling (\textit{yellow}).}
    \label{fig:phasediagram-triangular}
\end{figure}

\section{Qualitative phase diagram for the triangular lattice} \label{app:triangular}

To provide an overview of the different phases discussed in the main text, we present in Fig.~\ref{fig:phasediagram-triangular} a qualitative phase diagram of the triangular-lattice Rydberg atom array model. The phase boundaries were obtained from finite-PEPS calculations using open boundary conditions and bond dimension $D=4$, with selected calculations at $D=5$ yielding consistent results.

The phase diagram is not intended as a precise determination of the phase boundaries. Instead, it serves as a guide to the different ordering regimes. To faithfully accommodate the various crystalline density-wave patterns, different lattice sizes were employed depending on the expected ordering periodicity. In particular, lattice sizes between $22\times22$ and $25\times25$ were chosen such that the corresponding unit cells fit naturally into the finite system, thereby reducing finite-size effects arising from incommensurate boundaries. For example, phases with a $2\times2$ unit cell ($\frac13$- and $\frac14$-filling) were studied on $25\times25$ lattices, whereas phases exhibiting a $7\times7$ ($\frac17$-filling) periodicity were evaluated on $22\times22$ lattices.
The lattice sizes are chosen as an integer number of unit cells plus one additional site along each direction, such that the crystalline pattern terminates with an excitation at the boundary.

From the phase diagram, we can see that for low detuning $\Delta$, there is not enough incentive to have large fillings of Rydberg excitations, and the system stays trivially disordered. Increasing $\Delta$ favors higher excitation densities, while $R_b$ penalizes excitations that lie too close to each other. This competition leads to the emergence of density wave ordered states for larger detuning. This can be seen in the phase diagram, which features three distinct density wave ordered phases with excitation densities dictated by the value of $R_b$. Interestingly, there are two regions at $R_b \approx 1$ and $R_b \approx 1.9$ where the lattice geometry prohibits the stabilization of the desired excitation density, and the system remains trivially disordered, even for higher $\Delta$.

%%%%%%%%%%%%%%%%%%%%%%%%%%%%%%%%%%%%%%%%%%%%%%%%%%%%%%%%%%%%
\section{Kagomé lattice coarse-graining and boundary treatment} \label{app:kagome}

\begin{figure}[htbp]
	\centering
	\begin{subfigure}[b]{0.3\columnwidth}
		\centering
		\includegraphics[width=0.9\textwidth, page=1]{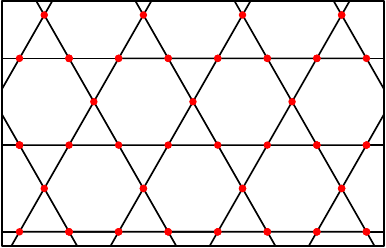}
		\caption%
		{{\small Kagomé Lattice}}    
		 
	\end{subfigure}
	\hfill
	\begin{subfigure}[b]{0.3\columnwidth}
		\centering
		\includegraphics[width=0.9\textwidth, page=2]{figures/KagomeCombineSites.pdf}
		\caption%
		{{\small Group 3 Sites together}}    
		\label{fig:kagometosquareb}
	\end{subfigure}
	\hfill
	\begin{subfigure}[b]{0.3\columnwidth}
		\centering
		\includegraphics[width=0.9\textwidth, page=3]{figures/KagomeCombineSites.pdf}
		\caption%
		{{\small Triangular Lattice}}    
		 
	\end{subfigure}
	\caption
	{\small To map the Kagomé lattice (a) onto the square lattice, we group three sites together and obtain a local physical dimension of $d = 2^3 = 8$ (b). The emerging structure is a triangular lattice (c), which can be mapped to the square lattice via the methods described in Sec~\ref{sec:results_triangular}.} 
	\label{fig:kagometosquare}
\end{figure}

The PEPS ansatz used in this work is formulated on a square lattice, while the Kagomé lattice has a different connectivity structure that cannot be represented directly without introducing artificial missing bonds. To overcome this mismatch, we coarse-grain the Kagomé lattice by combining three neighboring sites into a single effective site (see Fig.~\ref{fig:kagometosquare}), resulting in a triangular lattice that can subsequently be mapped onto the square lattice geometry, as done in App.~\ref{app:triangularmapping}.
Coarse-graining procedures of this type are standard in tensor-network approaches; see, e.g., \cite{schmoll_tensor_2023}, where six physical sites are grouped into a single tensor with local dimension $2^6 = 64$. 
The main advantage of this procedure is that the entanglement within each block is treated exactly, at the cost of an enlarged local Hilbert space.  

For finite systems with open boundary conditions, the Kagomé geometry also gives rise to dangling boundary sites that do not represent bulk physics and introduce spurious edge effects. To mitigate this, we project out these boundary degrees of freedom when computing bulk observables, ensuring that extracted quantities reflect the intrinsic behavior of the system rather than boundary artifacts.

\begin{figure}[htbp]
	\centering
	\begin{tikzpicture}[scale=1, every node/.style={draw, fill, circle, inner sep=1pt}]
		\foreach \x in {0,2,...,6}{
			\node[circle,draw,thick,fill=blue!50,minimum size=0.25cm] (A\x) at ({\x + 0.5}, {0}) {};
			\node[circle,draw,thick,fill=blue!50,minimum size=0.25cm] (B\x) at ({\x - 0.5}, {0}) {};
			\node[circle,draw,thick,fill=gray,minimum size=0.25cm] (C\x) at ({\x}, {sqrt(3)/2}) {};
			
			\draw[thick] (A\x) -- (B\x);
			\draw[thick] (B\x) -- (C\x);			
			\draw[thick] (A\x) -- (C\x);
			
			\draw[thick, dashed] (A\x) --++ (0.3, {-0.6*sqrt(3)/2});
			\draw[thick, dashed] (B\x) --++ (-0.3, {-0.6*sqrt(3)/2});
		}
	
		\draw[thick, dashed] (B0) --++ (-0.6, 0);
		\draw[thick, dashed] (A6) --++ (+0.6, 0);
		\draw[thick] (A0) -- (B2);
		\draw[thick] (A2) -- (B4);
		\draw[thick] (A4) -- (B6);
	\end{tikzpicture}
	\caption
	{\small The Kagomé lattice with open boundary conditions has dangling sites at the boundaries (gray) that could distort the ground-state nature because of these edge defects. To obtain less biased results, we project these sites out so that they do not influence the optimization.} 
	\label{fig:dangling}
\end{figure}
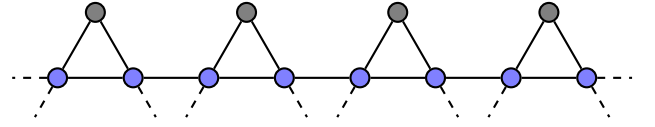

\section{Mapping from ruby to square lattice} \label{app:ruby}

\begin{figure}[htbp]
		\centering
		\begin{subfigure}[b]{0.49\columnwidth}
			\centering
			\resizebox{0.7\linewidth}{!}{\begin{tikzpicture}[scale=0.3, every node/.style={draw, fill, circle, inner sep=1pt}]
				\draw[thick] (-5,{-2*sqrt(3)}) -- (0,{3*sqrt(3)});
				\draw[thick] (5,{-2*sqrt(3)}) -- (0,{3*sqrt(3)});
				\draw[thick] (-5,{-2*sqrt(3)}) -- (5,{-2*sqrt(3)});
				\draw[thick] (-5,{2*sqrt(3)}) -- (5,{2*sqrt(3)});
				\draw[thick] (-5,0) -- (5,0);
				\draw[thick] (-2,-{3*sqrt(3)}) -- (4,{3*sqrt(3)});
				\draw[thick] (-5,0) -- (-2,{-3*sqrt(3)});
				\draw[thick] (2,{-3*sqrt(3)}) -- (-4,{3*sqrt(3)});
				\draw[thick] (2,{-3*sqrt(3)}) -- (5,0);
				\draw[thick] (-5,{2*sqrt(3)}) -- (-4,{3*sqrt(3)});
				\draw[thick] (5,{2*sqrt(3)}) -- (4,{3*sqrt(3)});

				\foreach \x in {-4,-2,0,2,4}{
					\node[circle,draw,thick,fill=blue!50,minimum size=0.2cm] at (\x,0) {};
					\node[circle,draw,thick,fill=blue!50,minimum size=0.2cm] at (\x,{2*sqrt(3)}) {};
					\node[circle,draw,thick,fill=blue!50,minimum size=0.2cm] at (\x,{-2*sqrt(3)}) {};
				}
				
				\foreach \x in {-4,4}{
					\node[circle,draw,thick,fill=blue!50,minimum size=0.2cm] at (\x,0) {};
					\node[circle,draw,thick,fill=blue!50,minimum size=0.2cm] at (\x,{-2*sqrt(3)}) {};
				}
				
				\foreach \x in {-4.5,-3.5,-0.5,0.5,3.5,4.5}{
					\node[circle,draw,thick,fill=blue!50,minimum size=0.2cm] at (\x,{-3/2*sqrt(3)}) {};
					\node[circle,draw,thick,fill=blue!50,minimum size=0.2cm] at (\x,{-1/2*sqrt(3)}) {};
				}
				
				\foreach \x in {-2.5,-1.5,1.5,2.5}{
					\node[circle,draw,thick,fill=blue!50,minimum size=0.2cm] at (\x,{3/2*sqrt(3)}) {};
					\node[circle,draw,thick,fill=blue!50,minimum size=0.2cm] at (\x,{1/2*sqrt(3)}) {};
					\node[circle,draw,thick,fill=blue!50,minimum size=0.2cm] at (\x,{-5/2*sqrt(3)}) {};
				}
				
				\node[circle,draw,thick,fill=blue!50,minimum size=0.2cm] at (-0.5,{5/2*sqrt(3)}) {};
				\node[circle,draw,thick,fill=blue!50,minimum size=0.2cm] at (0.5,{5/2*sqrt(3)}) {};
				\node[circle,draw,thick,fill=blue!50,minimum size=0.2cm] at (-3.5,{5/2*sqrt(3)}) {};
				\node[circle,draw,thick,fill=blue!50,minimum size=0.2cm] at (-4.5,{5/2*sqrt(3)}) {};
				\node[circle,draw,thick,fill=blue!50,minimum size=0.2cm] at (3.5,{5/2*sqrt(3)}) {};
				\node[circle,draw,thick,fill=blue!50,minimum size=0.2cm] at (4.5,{5/2*sqrt(3)}) {};
				
				\foreach \x in {-4,0,4}{
					\foreach \y in {-1.666,-0.333,2.333}{
						\draw[thick, red] (\x,{\y*sqrt(3)}) circle(1cm);
					}
				}
				\foreach \x in {-2,2}{
					\foreach \y in {-2.333,0.333,1.666}{
						\draw[thick, red] (\x,{\y*sqrt(3)}) circle(1cm);
					}
				}

			\end{tikzpicture}}
			\caption%
			{}    
			\label{fig:rubytosquareb}
		\end{subfigure}
		\begin{subfigure}[b]{0.49\columnwidth}
			\centering
			\resizebox{0.7\linewidth}{!}{\begin{tikzpicture}[scale=0.3, every node/.style={draw, fill, circle, inner sep=1pt}]
				
				\foreach \x in {-4,-2,0,2,4}{
					\draw[thick, dashed] (\x,{-2*sqrt(3)}) -- (\x,{2*sqrt(3)});
				}
				\foreach \y in {-2,0,2}{
					\draw[thick] (-4,{\y*sqrt(3)}) -- (4,{\y*sqrt(3)});
				}
				\foreach \x in {-4,0,4}{
					\draw[thick] (\x,{-2*sqrt(3)}) -- (\x,{0*sqrt(3)});
				}
				\foreach \x in {-2,2}{
					\draw[thick] (\x,{0*sqrt(3)}) -- (\x,{2*sqrt(3)});
				}
				\foreach \x in {-4,-2,0,2,4}{
					\foreach \y in {-2,0,2}{
						\node[circle,draw,thick,fill=red!50,minimum size=0.2cm] at (\x,{\y*sqrt(3)}) {};
					}
				}		
			\end{tikzpicture}}
			\caption%
			{}    
			\label{fig:rubytosquarec}
		\end{subfigure}
		\caption
		{\small Mapping from the ruby lattice (a), which is the link Kagomé lattice, to the square lattice (b) by coarse-graining $3$ atoms onto the same site. The dashed lines are additional bonds in our ansatz that are not present in the physical system.} 
		\label{fig:rubytosquare}
	\end{figure}
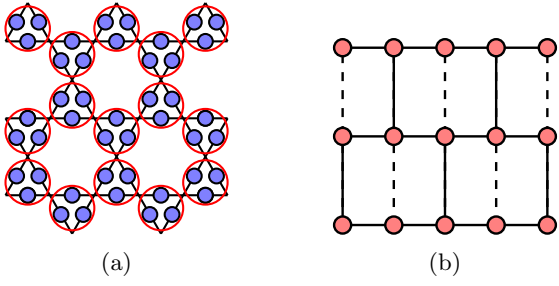

To study this lattice with finite PEPS algorithms, we again employ a coarse-graining scheme in which three sites of the ruby lattice are merged into a single effective site on a square lattice (Fig.~\ref{fig:rubytosquare}). 
This procedure increases the local Hilbert space dimension to $d = 2^3 = 8$ and modifies the effective connectivity. 
While each triangle in the ruby lattice is connected to three neighboring triangles, the corresponding sites on the square lattice exhibit four nearest neighbors. 
We test whether this modified connectivity introduces systematic biases and find no evidence for qualitative distortions in the resulting correlations, as can be seen from Table \ref{tab:mutualinformationruby}.

\begin{table}[h!]
    \centering
    \begin{tikzpicture}[
    scale=0.75,
    site/.style={draw, fill=blue!50, circle,
                 minimum size=0.18cm, inner sep=1pt},
    lab/.style={font=\scriptsize, inner sep=0pt}
]
    \node[site] (A) at (0,1) {$A$};
    \node[site] (B) at ({1.732050807568/2},1) {$B$};
    \node[site] (C) at ({1.732050807568},1) {$C$};
    \node[site] (D) at (0,0) {$D$};
    \node[site] (E) at ({1.732050807568/2},0) {$E$};
    \node[site] (F) at ({1.732050807568},0) {$F$};

    % Square-lattice bonds
    \draw[thick] (A)--(B)--(C)--(F)--(E)--(D)--(A);

    % Pairs of interest
    \draw[thick,dotted] (B)--(E);
\end{tikzpicture}

    \vspace{2pt}
    \caption{Mutual information for square-lattice-connected (disconnected) diagonal pairs $BE$ ($AF$ \& $CD$) across different parameter sets. Note that neither pair is physically connected in the Ruby lattice.}
    \renewcommand{\arraystretch}{1.4}
    \begin{tabular}{|c|c||c|c|c|}
        \hline
        \multicolumn{2}{|c||}{\textbf{Phase Parameters}}
        & \multicolumn{3}{c|}{\textbf{Mutual Information}} \\ \hline
        \textbf{Name} & $\Delta$ & $I_{BE}\,[10^{-3}]$
        & $I_{AF}\,[10^{-3}]$ & $I_{CD}\,[10^{-3}]$ \\ \hline
        Disordered & 1.3 & $1.06 \pm 0.05$
        & $1.01 \pm 0.06$ & $1.04 \pm 0.06$ \\ \hline
        Topological & 1.7 & $0.15 \pm 0.21$
        & $0.21 \pm 0.32$ & $0.37 \pm 0.61$ \\ \hline
        VBS & 3.0 & $36.1 \pm 52.1$
        & $38.4 \pm 41.8$ & $39.1 \pm 63.4$ \\ \hline
    \end{tabular}
    \label{tab:mutualinformationruby}
\end{table}

This suggests that the additional bonds of the square-lattice representation do not alter the essential physical structure of the original ruby lattice, and that the coarse-graining procedure provides a faithful effective description of the system.

 \section{Measurement of string operators and topological entanglement entropy} \label{app:topoEE_string_ops}

\begin{figure}[htbp]
	\centering
	\includegraphics[width=0.45\textwidth, page=5]{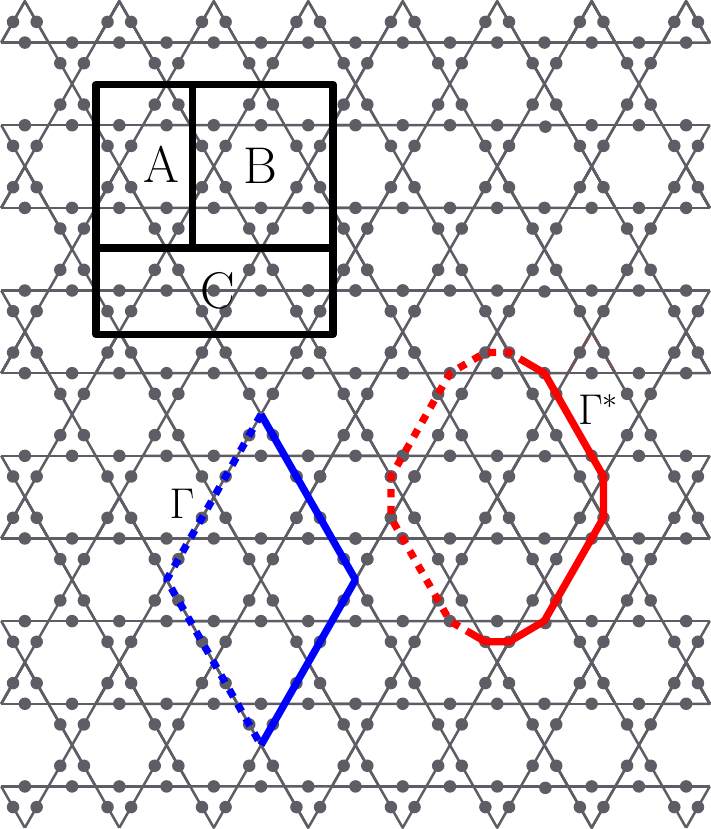}
	
	\caption
	{\small The different regions used to calculate the topological entanglement entropy via the Kitaev Preskill method ($A$, $B$ and $C$) and the paths used to evaluate the Fredenhagen–Marcu order parameter of the Wegner-Wilson lines (blue, $\Gamma$) and the 't Hooft lines (red, $\Gamma^*$). The solid and dashed segments are treated as independent open strings, while their union forms the corresponding closed loop used for the normalization. The sizes and shapes of the regions and paths match those used in the calculations, but their positions have been shifted for graphical clarity.} 
	\label{fig:topoEE_WT_paths}
\end{figure}

To calculate the topological correction to the area law of entanglement entropy, we employ the Kitaev-Preskill method, 
\begin{equation}
    \gamma = S_{AB} + S_{BC} + S_{AC} - S_{A} - S_{B} - S_{C} - S_{ABC},
\end{equation}
with the partitions shown in Fig.~\ref{fig:topoEE_WT_paths}. In principle, larger subsystems are desirable, as this expression becomes exact only in the thermodynamic limit when the subsystem boundaries are much larger than the correlation length. In practice, however, the statistical uncertainty of the swap-trick estimator increases rapidly with subsystem size, limiting the accessible region sizes.

To compute the Fredenhagen-Marcu order parameters, we used the paths shown in Fig.~\ref{fig:topoEE_WT_paths}. Each colored path consists of a solid and a dashed segment. The solid and dashed segments are treated as independent open strings, while their union defines the corresponding closed loop entering the normalization. To improve statistics, we evaluate the order parameter for multiple translationally equivalent copies of each path on the lattice and report the average value.

%%%%%%%%%%%%%%%%%%%%%%%%%%%%%%%%%%%%%%%%%%%%%%%%%%%%%%%%%%%

\section{Tensor-network construction and analytical understanding of the topological phase} \label{app:one_parameter_ansatz}
\begin{figure}[htbp]
\centering

%\begin{minipage}[htbp]{0.48\linewidth}
\centering
	\begin{tikzpicture}[scale=0.8]
		
		\draw[gray!50,thick] ({-0.5*sqrt(3)}, 3.5) -- ({4.5*sqrt(3)},3.5);
		\draw[gray!50,thick] ({-0.5*sqrt(3)}, 0.5) -- ({4.5*sqrt(3)},0.5);
		\draw[gray!50,thick] ({0*sqrt(3)}, 5) -- ({2*sqrt(3)},-1);
		\draw[gray!50,thick] ({2*sqrt(3)}, 5) -- ({4*sqrt(3)},-1);
		\draw[gray!50,thick] ({0*sqrt(3)}, -1) -- ({2*sqrt(3)},5);
		\draw[gray!50,thick] ({2*sqrt(3)}, -1) -- ({4*sqrt(3)},5);

		\foreach \x in {0,2,4}{
			\foreach \y in {0,4}{
				\draw[thick] ({\x * sqrt(3) -0.25},{\y - 0.25}) rectangle ({\x*sqrt(3)+0.25},{\y + 0.25});
				\node (t\x\y) at ({\x * sqrt(3)},{\y}) {$t$};
                \draw[thick] (t\x\y) -- ++(0.35, 0.60621);
		}	
		}
		\foreach \x in {1,3}{
			\foreach \y in {1,3}{
				\draw[thick] ({\x * sqrt(3) -0.25},{\y - 0.25}) rectangle ({\x*sqrt(3)+0.25},{\y + 0.25});
				\node (t\x\y) at ({\x * sqrt(3)},{\y}) {$t$};
                \draw[thick] (t\x\y) -- ++(0.35, 0.60621);
			}	
		}
	
		\draw[thick] (t00)--(t11);
		\draw[thick] (t11)--(t13);
		\draw[thick] (t04)--(t13);
		\draw[thick] (t24)--(t13);	
		\draw[thick] (t24)--(t33);	
		\draw[thick] (t33)--(t31);	
		\draw[thick] (t11)--(t20);	
		\draw[thick] (t31)--(t20);	
		\draw[thick] (t31)--(t40);	
		\draw[thick] (t33)--(t44);

		\node[circle, draw, fill=white, inner sep=2pt] at ({0.5*sqrt(3)},0.5) {$T$};
		\node[circle, draw, fill=white, inner sep=2pt] at ({1.5*sqrt(3)},0.5) {$T$};
		\node[circle, draw, fill=white, inner sep=2pt] at ({2.5*sqrt(3)},0.5) {$T$};
		\node[circle, draw, fill=white, inner sep=2pt] at ({3.5*sqrt(3)},0.5) {$T$};
		\node[circle, draw, fill=white, inner sep=2pt] at ({0.5*sqrt(3)},3.5) {$T$};
		\node[circle, draw, fill=white, inner sep=2pt] at ({1.5*sqrt(3)},3.5) {$T$};
		\node[circle, draw, fill=white, inner sep=2pt] at ({2.5*sqrt(3)},3.5) {$T$};
		\node[circle, draw, fill=white, inner sep=2pt] at ({3.5*sqrt(3)},3.5) {$T$};
		\node[circle, draw, fill=white, inner sep=2pt] at ({1*sqrt(3)},2) {$T$};
		\node[circle, draw, fill=white, inner sep=2pt] at ({3*sqrt(3)},2) {$T$};
		
		\foreach \x in {0,2,4}{
			\foreach \y in {0,4}{
				\draw[thick, fill=white] ({\x * sqrt(3) -0.25},{\y - 0.25}) rectangle ({\x*sqrt(3)+0.25},{\y + 0.25});
				\node (t\x\y) at ({\x * sqrt(3)},{\y}) {$t$};
			}	
		}
		\foreach \x in {1,3}{
			\foreach \y in {1,3}{
				\draw[thick, fill=white] ({\x * sqrt(3) -0.25},{\y - 0.25}) rectangle ({\x*sqrt(3)+0.25},{\y + 0.25});
				\node (t\x\y) at ({\x * sqrt(3)},{\y}) {$t$};
			}	
		}
	
	\end{tikzpicture}
 \caption{{\small Schematic drawing of the toric code inspired wave function as a bond dimension $2$ PEPS. The tensors $T$ sit on the sites of the Kagomé lattice, and the tensors $t$ sit inside the triangle, representing $3$ Rydberg atoms each. Note that the $t$ tensors have a physical index as well, which is not depicted. In gray, we show the underlying Kagomé lattice, and in black, the connections between the tensors.}}
\label{fig:one_p_ansatz}
\end{figure}
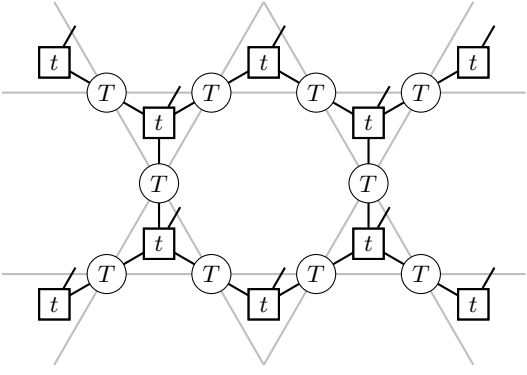

To gain analytical insight into the topologically ordered regime of the PXP model, we construct a minimal PEPS inspired by the exact bond-dimension-2 PEPS representation of the toric-code ground state \cite{verstraete_criticality_2006, orus_practical_2014}. 
To connect this construction to the PXP model on the Ruby lattice, we observe that Ruby lattice sites correspond to links of the Kagomé lattice. A Rydberg excitation on a Ruby site can thus be viewed as a dimer occupying the corresponding Kagomé link.
The three Ruby spins associated with each Kagomé triangle are encoded in a tensor ($t$) placed at its center, whose physical index ($s$) labels the four allowed configurations: no excitation $s=0$: \begin{tikzpicture}[scale=0.4]
	\draw[thick] (0,0) -- (1,0) -- (0.5, {sqrt(3)/2}) -- (0,0);
\end{tikzpicture} or a single excitation on one of the three links $s=1$: \begin{tikzpicture}[scale=0.4]
\draw[thick] (0,0) -- (1,0) -- (0.5, {sqrt(3)/2}) -- (0,0);
\draw[thick, fill=red!50] (0.5,0) ellipse (0.5cm and 0.1cm);
\end{tikzpicture}, $s=2$: \begin{tikzpicture}[scale=0.4]
\draw[thick] (0,0) -- (1,0) -- (0.5, {sqrt(3)/2}) -- (0,0);
\draw[thick, fill=red!50, rotate around={60:(0,0)}] (0.5,0) ellipse (0.5cm and 0.1cm);
\end{tikzpicture}, $s=3$: \begin{tikzpicture}[scale=0.4]
\draw[thick] (0,0) -- (1,0) -- (0.5, {sqrt(3)/2}) -- (0,0);
\draw[thick, fill=red!50, rotate around={-60:(1,0)}] (0.5,0) ellipse (0.5cm and 0.1cm);
\end{tikzpicture}. In principle, the local Hilbert space dimension should be $d=8$, but in the PXP limit four of those are forbidden because they contain double excitations and can be omitted.
We place $T$ tensors on every site of the Kagomé lattice to enforce the dimer constraint. 
This leads to a PEPS with bond dimension $2$ (see Fig.~\ref{fig:one_p_ansatz}) with tensor entries specified in Eq.~\ref{eq:bigT} and Eq.~\ref{eq:smallt}.

%\end{minipage}
%\hfill

\begin{align}
\label{eq:bigT}
T_{i_1,i_2} &=
\begin{cases}
1 & \text{, for } i_1+i_2 = 1 \mod 2\\
0 & \text{, otherwise}
\end{cases}
\\
\label{eq:smallt}
t_{j_1,j_2,j_3}^s &=
\begin{cases}
1 & \text{, for } j_1=j_2=j_3=s=0\\
\sqrt{1-p^2} & \text{, for } j_k=0,\ j_1+j_2+j_3=2,\ s=k\\
-p & \text{, for } j_k=0,\ j_1+j_2+j_3=2,\ s=0\\
0 & \text{, otherwise}
\end{cases}
\end{align}
 
In order to incorporate $e$-anyon fluctuations needed to drive resonances with a single-site Rabi term, we have to relax the maximal dimer filling constraint and introduce the parameter $p$. 
This parameter controls the amplitude for configurations in which a Rydberg excitation is omitted despite being allowed by the surrounding $T$ tensors. 
For $p=0$, only maximally packed dimer coverings are allowed.

The ansatz drastically reduces the number of independent variational parameters from $\mathcal{O}(L_x L_y D^4 d)$ (all entries of the PEPS vary freely) to just a single one, namely $p$. 
Optimizing $p$ for several points within the spin-liquid regime yields ground-state energies that closely match those obtained from unrestricted $D=5$ PEPS optimizations; see Tab.~\ref{tab:energy-comparison} for a comparison. 

\begin{table}[!htbp]
    \vspace{0.5cm}  
    \centering
    \renewcommand{\arraystretch}{1.4}
    \begin{tabular}{|c||c|c||c|}
        \hline
        $\Delta$ & $p$ & $\langle \hat{H} \rangle_p$ & $\langle \hat{H} \rangle$ \\
        \hline
        1.7 & 0.2707 & $-0.49427 \pm 0.00005$ & $-0.4947 \pm 0.0005$\\ \hline
        1.8 & 0.2595 & $-0.519000 \pm 0.000006$ & $-0.5199 \pm 0.0003$\\ \hline
        1.9 & 0.2473 & $-0.544260 \pm 0.000006$ & $-0.5443 \pm 0.0005$\\ \hline
        2.0 & 0.2363 & $-0.569656 \pm 0.000004$ & $-0.5690 \pm 0.0007$\\ \hline
    \end{tabular}
    \caption{Comparison of the variational energy obtained from the one-parameter PEPS ansatz, $\langle\hat{H}\rangle_p$, and the fully optimized PEPS energy, $\langle\hat{H}\rangle$, for different detunings $\Delta$.}
    \label{tab:energy-comparison}
\end{table}

\begin{figure}[H]
    \vspace{0.5cm}  
\centering
 \includegraphics[width=0.7\linewidth]{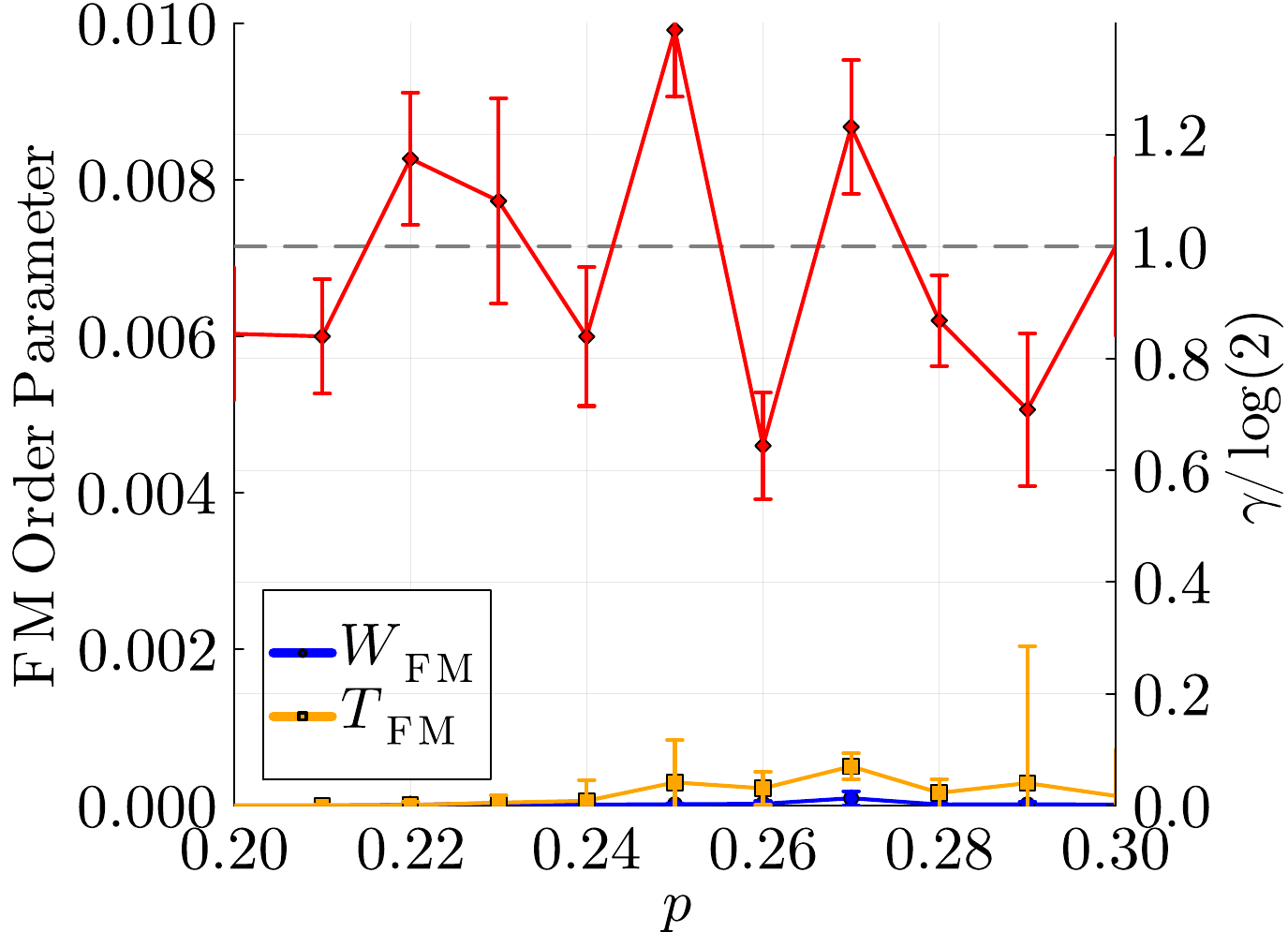}
\caption{{\small Topological entanglement entropy (red diamonds) and the Fredenhagen-Marcu order parameter (blue circles and orange squares) for one-parameter PEPS with different values of $p$. Consistent with a topologically ordered state, the FM order parameters approach zero, while the topological entanglement entropy fluctuates around the expected value of $\gamma = \log 2$ (gray).}}
\label{fig:FMOPS_p}
\end{figure}

At $p=0$, the ansatz can be shown analytically to possess $\mathbb{Z}_2$ topological order. Furthermore, this result is stable to first order in $p$. Numerical calculations for the optimized values of $p$ yield a topological entanglement entropy close to $\log 2$ together with vanishing Fredenhagen–Marcu order parameters Fig.~\ref{fig:FMOPS_p}. Taken together, the topological diagnostics and the close agreement with unrestricted PEPS energies indicate that the simple ansatz captures the essential physics of the phase. The preceding observations are consistent with a finite-size topological phase. Whether this topological order survives in the thermodynamic limit, however, remains an open question.

\end{document}